\documentclass[%
 reprint,
superscriptaddress,
nofootinbib,
 amsmath,amssymb,
 aps,
 prd,
]{revtex4-2}

\usepackage[utf8]{inputenc}
\DeclareUnicodeCharacter{2009}{\,}
\DeclareUnicodeCharacter{0229}{\,} 
\usepackage{amsmath}
\usepackage{amssymb}
\usepackage{mathtools}
\usepackage{xcolor}
\usepackage{graphicx}
\usepackage{listings}\usepackage{color}
\usepackage{hyperref}
\hypersetup{
    colorlinks=true,
    linkcolor=blue,
    filecolor=magenta,      
    urlcolor=blue,
    citecolor=violet
}

\usepackage{dcolumn}
\usepackage{multirow}

\definecolor{dkgreen}{rgb}{0,0.6,0}
\definecolor{gray}{rgb}{0.5,0.5,0.5}
\definecolor{mauve}{rgb}{0.58,0,0.82}

\def\fpeak{f_{\mathrm{peak}}}
\def\fpi{f_{\mathrm{peak},i}}
\def\fpit{f_{\mathrm{peak},i,\mathrm{true}}}
\def\hfp{\hat{f}_{\mathrm{peak}}}

\def\hfpij{\hat{f}_{\mathrm{peak},ij}}
\def\mc{\mathcal{M}}
\def\mci{\mathcal{M}_i}
\def\mcit{\mathcal{M}_{i,\mathrm{true}}}
\def\R{R_{1.6}}
\def\msun{\mathrm{M_{\odot}}}

\newcommand{\emdashb}{\textbf{---} }

\begin{document}
\title{Needle in a Bayes Stack: a Hierarchical Bayesian Method for Constraining the Neutron Star Equation of State with an Ensemble of Binary Neutron Star Post-merger Remnants}

\author{Alexander W. Criswell}\email{crisw015@umn.edu}
\affiliation{Minnesota Institute for Astrophysics, University of Minnesota, Minneapolis, Minnesota 55455, USA}
% \correspondingauthor{Alexander Criswell}

\author{Jesse Miller}\affiliation{School of Statistics, University of Minnesota, Minneapolis, Minnesota 55455, USA}
\author{Noah Woldemariam}
\affiliation{Department of Computer Science and Engineering, University of Minnesota, Minneapolis, Minnesota 55455, USA}
\author{Theodoros Soultanis}
\affiliation{GSI  Helmholtzzentrum  f\"ur  Schwerionenforschung,  Planckstra{\ss}e  1,  64291  Darmstadt,  Germany}
\affiliation{Heidelberg Institute for Theoretical Studies, Schloss-Wolfsbrunnenweg 35,
69118 Heidelberg, Germany}
\affiliation{Max-Planck-Institut f\"ur Astronomie, K\"onigstuhl 17, 69117 Heidelberg, Germany}
\author{Andreas Bauswein}
\affiliation{GSI  Helmholtzzentrum  f\"ur  Schwerionenforschung,  Planckstra{\ss}e  1,  64291  Darmstadt,  Germany}
\affiliation{Helmholtz Research Academy Hesse for FAIR (HFHF), GSI Helmholtz Center for Heavy Ion Research, Campus Darmstadt, 64291 Darmstadt, Germany}
\author{Katerina Chatziioannou}
\affiliation{Department of Physics, California Institute of Technology, Pasadena, California 91125, USA}
\affiliation{LIGO Laboratory, California Institute of Technology, Pasadena, California 91125, USA}
\author{Michael W. Coughlin}
\affiliation{Minnesota Institute for Astrophysics, University of Minnesota, Minneapolis, Minnesota 55455, USA}
\author{Galin Jones}
\affiliation{School of Statistics, University of Minnesota, Minneapolis, Minnesota 55455, USA}
\author{Vuk Mandic}
\affiliation{Minnesota Institute for Astrophysics, University of Minnesota, Minneapolis, Minnesota 55455, USA}

\date{\today}

\begin{abstract}
Binary neutron star (BNS) post-merger gravitational-wave emission can occur in the aftermath of a BNS merger \emdashb provided the system avoids prompt collapse to a black hole \emdashb as a quasistable hypermassive remnant experiences quadrupolar oscillations and non-axisymmetric deformations. The post-merger gravitational-wave spectrum possesses a characteristic peak frequency that has been shown to be dependent on the binary chirp mass and the neutron star equation of state (EoS), rendering post-merger gravitational waves a powerful tool for constraining neutron star composition. Unfortunately, the BNS post-merger signal is emitted at high ($\gtrsim 1.5$ kHz) frequencies, where ground-based gravitational wave detectors suffer from reduced sensitivity. It is therefore unlikely that post-merger signals will be detected with sufficient signal-to-noise ratio (SNR) until the advent of next-generation detectors. However, by employing empirical relations derived from numerical relativity simulations, we can combine information across an ensemble of BNS mergers, allowing us to obtain EoS constraints with many low-SNR signals. We present a hierarchical Bayesian method for deriving constraints on $\R$, the radius of a 1.6$\msun$ neutron star, through an ensemble analysis of binary neutron star mergers. We apply this method to simulations of the next two LIGO-Virgo-KAGRA observing runs, O4 and O5, as well as an extended 4-year run at A+ sensitivity, demonstrating the potential of our approach to yield EoS information from the post-merger signal with current-generation detectors. The A+ 4-year scenario is predicted to improve the constraint on $\R$ from the currently available multimessenger-based 95\% credible interval (C.I.) uncertainty of $\R=12.07^{+0.98}_{-0.77}$ km to $\R=11.91^{+0.80}_{-0.56}$ km, a 22\% reduction of the 95\% C.I. width.
\end{abstract}

\maketitle

\section{Introduction}

\subsection{On Neutron Star Interiors}

The nature of matter at the temperatures and densities present in neutron star (NS) interiors is an open question. This is, in some ways, unsurprising; NSs are among the most extreme material objects in the universe. Highly relativistic and immensely dense, with masses/radii of order 1-2 $\msun$/10-15 km, NSs reside in the final stable bastion of matter before collapse into a black hole and are the sole known occupants of the cold, high-density regime of dense nuclear matter \citep{baym_hadrons_2018}. As such, NSs provide a powerful tool for exploring dense nuclear theory. 

In practice, one wishes to constrain the NS equation of state (EoS), which describes pressure as a function of mass-energy density (or, equivalently, mass as a function of radius). Such constraints are realized through a variety of NS observables, including direct mass-radius measurements of individual pulsars by NICER (e.g., \citep{riley_nicer_2019,miller_PSR_2019,riley_nicer_2021,miller_radius_2021}), observations of individual high-mass NSs \citep{demorest_two-solar-mass_2010,antoniadis_massive_2013,cromartie_relativistic_2020,fonseca_refined_2021,romani_psr_2022}, measurements of the NS tidal deformability $\Lambda$ from binary neutron star (BNS) inspiral gravitational wave (GW) signals by the LIGO-Virgo-KAGRA (LVK) detectors \citep{abbott_gw170817_2017,abbott_gw170817_2018,abbott_gw190425_2020}, and studies of the kilonova (AT2017gfo) \citep{abbott_multi-messenger_2017,soares-santos_electromagnetic_2017,cowperthwaite_electromagnetic_2017,nicholl_electromagnetic_2017,chornock_electromagnetic_2017,margutti_electromagnetic_2017,alexander_electromagnetic_2017} and gamma ray burst (GRB170817) \citep{abbott_multi-messenger_2017,goldstein_ordinary_2017,savchenko_integral_2017} counterparts to the sole BNS merger detected in both GWs and electromagnetic radiation, GW170817. By combining the information contained in each of these measurements in a cohesive multimessenger framework, one can arrive at the most stringent available constraints. At the time of writing, a number of complementary studies have presented such collective constraints, reaching similar conclusions: a (95\% credible interval (C.I.)) constraint on the radius of a $1.6\msun$ NS of  $\R=12.07^{+0.98}_{-0.77}$ km \citep{coughlin_multi-messenger_2019,raaijmakers_constraining_2020,miller_PSR_2019,dietrich_multimessenger_2020,landry_nonparametric_2020,capano_stringent_2020,raaijmakers_constraints_2021,al-mamun_combining_2021,biswas_impact_2021,pang_nuclear-physics_2021,breschi_at2017gfo_2021,miller_radius_2021,legred_impact_2021,nicholl_tight_2021,huth_constraining_2022}.

\subsection{Binary Neutron Star Mergers and Their Remnants}
BNS mergers consist of two phases, both of which produce GWs: inspiral and post-merger. The inspiral signal encodes a wealth of information about the merging binary, including the binary masses, spins, distance, redshift, tidal deformation, and so on, although some of these parameters are degenerate with each other. Importantly for this work, the inspiral frequency evolution is dependent (to leading order) on a specific combination of the binary masses, known as the chirp mass:
\begin{equation}
\mc = \frac{(m_1 m_2)^{3/5}}{(m_1 + m_2)^{1/5}}\,,
\end{equation}\label{chirpmassdef}
where $m_1$ and $m_2$ are the individual NS masses. The majority of the frequency content and frequency evolution of a GW binary inspiral signal occurs in the most sensitive band of the LIGO \citep{aasi_advanced_2015}, Virgo \citep{acernese_advanced_2015}, and KAGRA \citep{somiya_detector_2012,aso_interferometer_2013} detectors, allowing inspiral analyses to measure the chirp mass $\mc$ to extremely high precision. For example, the chirp mass of GW170817 was measured to be $1.188^{+0.004}_{-0.002} \msun$ (90\% C.I.) \citep{abbott_gw170817_2017}.

After the inspiral phase, there are four possible outcomes for the post-merger remnant object: direct collapse to a black hole, the formation of a short-lived hypermassive neutron star (HMNS), the formation of a long-lived supramassive neutron star (SMNS), and the formation of a stable NS. Given our understanding of the BNS mass distribution, it is likely that a large fraction of BNS events will result in the second scenario: formation of a HMNS that collapses to a black hole within a few seconds \citep{baumgarte_maximum_1999,hotokezaka_remnant_2013}. Indeed, it has been argued that this was the nature of GW170817's remnant (e.g., \citep{margalit_constraining_2017,granot_lessons_2017,perego_at_2017,ma_gw170817_2018,ruiz_gw170817_2018,pooley_gw170817_2018,rezzolla_using_2018,metzger_magnetar_2018,gill_when_2019}.) This scenario is of particular interest to us, as it is accompanied by a post-merger GW signal with frequencies of 1.5-4 kHz and properties that are strongly tied to the NS EoS. 

HMNSs exceed the maximum allowed mass of a cold, uniformly-rotating NS;\footnote{Thus surpassing the limitations of the SMNS scenario, which exceeds the maximum allowed mass of a cold, non-spinning NS through some degree of uniform rotation.} such a configuration should, in principle, be unstable and undergo gravitational collapse. However, as shown in \citet{baumgarte_maximum_1999}, a differentially-rotating NS featuring a higher rotation rate in its core than its envelope \emdashb  of the kind formed in many BNS mergers \citep{shibata_simulation_2000} \emdashb can temporarily maintain support for much higher\footnote{An increase of 30-70\%; see \citet{bauswein_prompt_2013}.} masses against gravitational collapse into a black hole. This support can be compromised by neutrino losses, gravitational-wave emission, and angular momentum transport within the NS reducing the degree of differential rotation present \citep{shapiro_differential_2000,hotokezaka_remnant_2013}. Precisely modelling these quenching processes is challenging, and as such the exact timescale for loss of support from differential rotation is difficult to specify. However, it is generally agreed that this process happens on relatively short timescales \citep{shapiro_differential_2000,hotokezaka_remnant_2013}.

\begin{figure}
    \centering
    \includegraphics[width=1\linewidth]{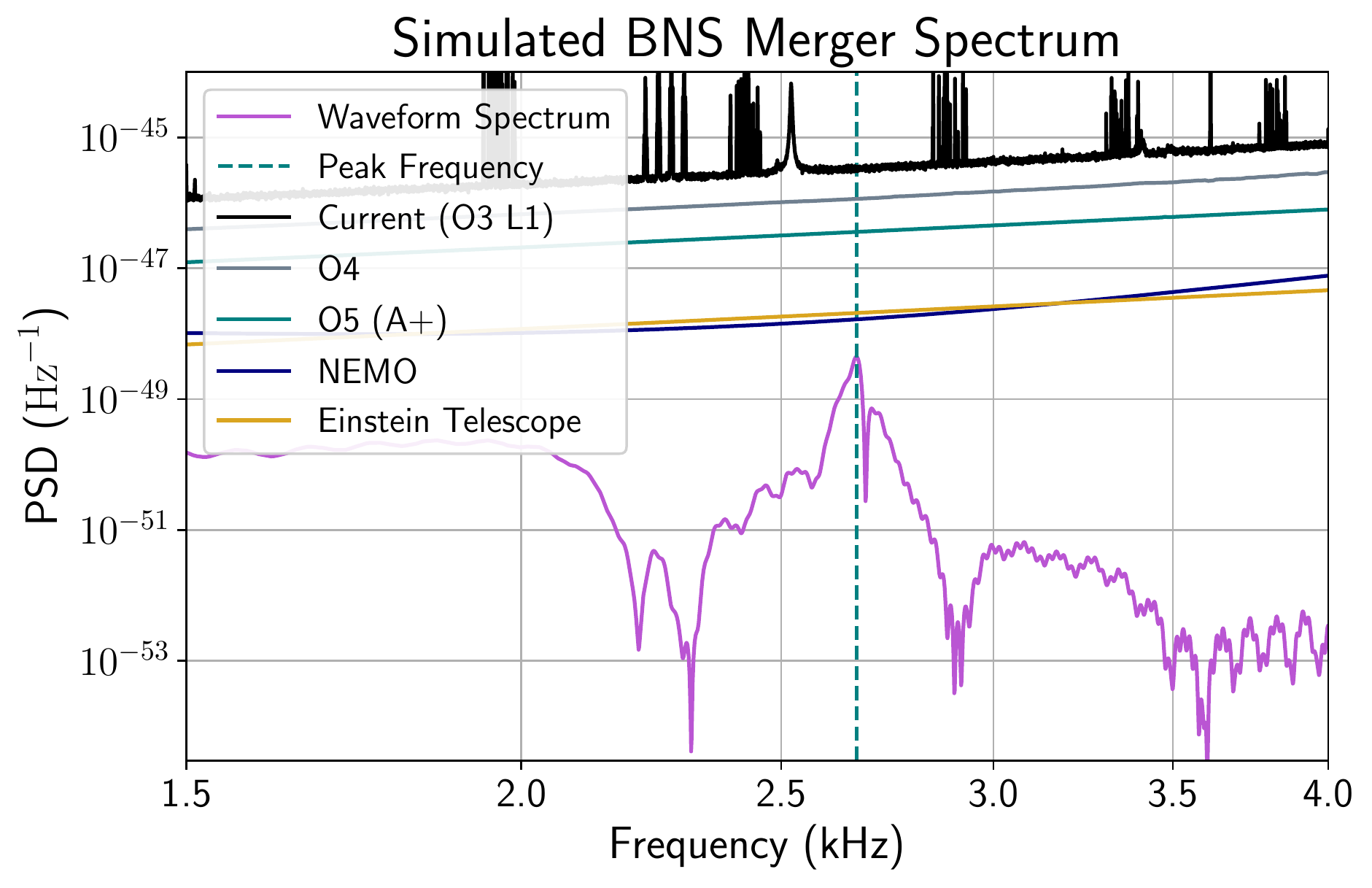}
    \caption{GW power spectral density (PSD) of a simulated merger + post-merger waveform for the SFHX EoS (see \S \ref{waveforms}) with $m_1 = 1.03\msun$ and $m_2 = 1.09\msun$, injected at a distance of 50 Mpc. The current, projected, and proposed detector noise PSDs discussed in this work are provided for reference: that of the LIGO Livingston detector during the 3rd LVK observing run (O3 L1); the projected LIGO sensitivities for the 4th and 5th LVK observing runs (O4 and O5), assuming LIGO A+ design sensitivity for O5 \citep{oreilly_ligo-t2000012-v1_2020}; and the proposed NEMO and Einstein Telescope detectors (see \S~\ref{future-detection}) \citep{evans_ligo-t1500293-v13_2020,adya_ligo-t2000062-v1_2020}. The post-merger peak frequency of 2.67 kHz is marked.}
    \label{fig:PM-spectrum}
\end{figure}

Numerical relativity simulations of hypermassive BNS remnants have shown that the dominant frequency of GW emission in such remnants is driven by $f$-mode pulsations\footnote{The spherical harmonic $l=m=2$ mode of oscillation. See \citet{stergioulas_gravitational_2011a} for details concerning the generation of GWs in post-merger remnants through $f$-mode oscillations and non-axisymmetric deformations. A review of spherical harmonic modes as applied to asteroseismology of compact objects can be found in \citet{stergioulas_rotating_2003}.} and the rotating bar-shaped structure of the remnant that, due to their quadrupolar nature, are efficient GW emitters; additionally, the post-merger spectrum possesses a number of subdominant peaks arising from other oscillation modes or non-linear dynamical features \citep{zhuge_gravitational_1994, shibata_merger_2005,stergioulas_gravitational_2011a,takami_spectral_2015,bauswein_unified_2015,soultanis_analytic_2022}. We follow the convention of referring to this characteristic dominant frequency as the post-merger peak frequency $\fpeak$.\footnote{Sometimes referred to elsewhere as $f_2$.} In Fig. \ref{fig:PM-spectrum}, we plot the spectrum of a simulated BNS gravitational waveform, drawing attention to the post-merger peak frequency. 

Interestingly, $\fpeak$ has been shown to be strongly dependent on the NS EoS and, for fixed mass, shows significant correlations with a number of EoS-dependent quantities. This EoS dependence has been for instance described in terms of the tidal coupling constant $\kappa_2^T$ \citep{bernuzzi_modeling_2015,takami_spectral_2015} and the radii of cold, non-rotating NS of a fixed mass (e.g., that of a $1.6\msun$ NS, $\R$) \citep{bauswein_equation--state_2012,bauswein_measuring_2012}. These correlations imply that the post-merger GW signal (or, at least, its peak frequency) is primarily determined by the merger mass and the NS EoS, and can serve as a powerful tool for constraining the nuclear EoS. Additionally, if EoS constraints arising from post-merger studies differ significantly from those arising from inspiral measurements or other studies of cold, stable NSs, this could be indicative of phase transitions occurring during the merger process and provide evidence for more exotic matter in NS cores such as deconfined quark matter \citep{bauswein_identifying_2019}. 

Unfortunately, the post-merger GW signal is predicted to be difficult to detect \citep{clark_prospects_2014,torres-rivas_observing_2019-1}; while its total emitted power is in fact larger than that of the inspiral signal, the post-merger signal is emitted at high (1.5-4 kHz) frequencies, where the LVK detectors are less sensitive. Unambiguous individual detection of post-merger GW by current-generation detectors is unlikely unless a suitable BNS merger occurs within $\lesssim 20$ Mpc, corresponding to a post-merger signal-to-noise ratio (SNR)$~\gtrsim4-8$ \citep{clark_prospects_2014,tsang_modeling_2019}.

Indeed, while there have been two detected BNS mergers in the first three LIGO/Virgo observing runs, no post-merger GW emission has been detected. For GW170817, searches were performed for post-merger GWs of short ($\lesssim$ 1 s - the variety considered in this work), intermediate ($\lesssim$ 500 s), and long ($\lesssim$ 8.5 days) durations.\footnote{We limit this discussion to searches for short-duration signals from HMNS remnants; see \citet{abbott_search_2017} and \citet{abbott_search_2019} for details on the searches for intermediate- and long-duration signals, respectively.} The search for short-duration signals employed two primary approaches. The first was a maximum-likelihood evaluation of coherent excess power using the Coherent WaveBurst (cWB) algorithm \citep{klimenko_method_2016}. The second was a BayesWave search of the kind outlined in \citet{chatziioannou_inferring_2017} (see \S\ref{bayeswave}). Neither search yielded a detection of the post-merger signal;  the cWB search placed upper limits on the root-sum-square GW strain from 1-4 kHz of $h_{rss}^{50\%} = 8.4\times10^{-22}\text{ Hz}^{-1/2}$ \citep{abbott_search_2017} and the BayesWave analysis placed 90\% credible upper limits on the network SNR from 1-4 kHz of $\rho_{\text{net}} \le 6.7$, as well as frequency-dependent 90\% upper limits on the strain amplitude spectral density and the spectral energy density \citep{abbott_properties_2019}. These upper limits are larger than any expected strain produced by the post-merger remnant of a GW170817-like event; the LIGO/Virgo high-frequency sensitivity was simply insufficient to detect the post-merger signal \citep{abbott_search_2017,abbott_properties_2019}.

For GW190425, only a BayesWave search was performed. No post-merger signal was detected and 90\% credible upper limits were placed on the strain amplitude spectral density of $1.1\times10^{-22}\text{ Hz}^{-1/2}$ for a frequency of 2.5 kHz, higher than those placed for GW170817 due to increased source distance \citep{abbott_gw190425_2020}. It is worth noting in this case that not only is this strain upper limit higher than would be expected for a post-merger signal, but the inferred binary masses for GW190425 are so large that the likely post-merger scenario for this binary is direct collapse to a black hole (and therefore no post-merger GW signal emitted in the first place) \citep{bauswein_prompt_2013,agathos_inferring_2020,abbott_gw190425_2020}.

Taken together, this indicates that unambiguous individual post-merger detections will need to wait until next-generation detectors \citep{torres-rivas_observing_2019-1}. With recent improvements to analytic post-merger models \citep{breschi_kilohertz_2019,breschi_kilohertz_2022-1,easter_detection_2020,soultanis_analytic_2022}, Einstein Telescope and Cosmic Explorer are predicted to observe $\lesssim 2-4$ and $\lesssim 10$ post-merger signals with sufficient SNR per year, respectively \citep{martynov_exploring_2019,breschi_kilohertz_2022}; these rates can be elevated to $\lesssim 40-120$ if one considers a current + next-generation network, multiple next-generation detectors, or tuning Cosmic Explorer to optimally detect the post-merger signal \citep{evans_horizon_2021,srivastava_science-driven_2022}. Given that the prospects for unambiguous individual detection of post-merger GWs in the near future seem bleak \emdashb and thus their constraining power for the NS EoS appears, for now, out of reach \emdashb it is worth considering how we might make use of the ambiguous and marginal detections we \textit{may} be able to achieve with current-generation detectors. 

One promising approach is to combine many such signals in a coherent fashion, allowing us to learn what the ensemble of BNS mergers can tell us in concert, rather than individually, about the NS EoS. Previous studies have proposed coherent mode stacking using empirical relations and inspiral mass measurements \citep{bose_neutron-star_2018} or phase information from the inspiral signal \citep{yang_gravitational_2018}, although this latter method is predicted to only produce meaningful results for next-generation detectors. There remains, however, another potential tactic for aggregating information about the NS EoS across many marginal post-merger signals: that of hierarchical Bayesian inference.

\subsection{Hierarchical Bayesian Inference}
Bayesian inference is a common framework for parameter estimation in GW astronomy. In its most basic application, we wish to calculate $p(\vec{\theta}|d)$, the posterior probability density of a set of model parameters $\vec{\theta} = \langle\theta_1...\theta_M\rangle$, given some GW data $d$. We use Bayes' theorem to do so:
\begin{equation}
    p(\vec{\theta}|d) = \frac{p(d|\vec\theta)}{p(d)}p(\vec\theta) \propto p(d|\vec\theta)p(\vec\theta),
\end{equation}
where $p(d|\vec\theta)$ is the likelihood function of the data given a signal model parameterized by $\vec\theta$, $p(d)$ is the evidence, and $p(\vec\theta)$ is the prior probability of the parameters $\vec\theta$. If we have multiple independent events $i = 1...N$, the joint likelihood is
\begin{equation}
    p(d_1 ... d_N|\vec{\theta}) = \prod_{i=1}^N p(d_i|\vec{\theta}),
\end{equation}
and the posterior is
\begin{equation}
\begin{split}
    p(\vec{\theta}|d_1 ... d_N) & =  \frac{1}{p(d)} \prod_{i=1}^N \left[p(d_i|\vec{\theta})\right]p(\vec{\theta})\\
    & \propto \prod_{i=1}^N \left[p(d_i|\vec{\theta})\right]p(\vec{\theta}),
\end{split}
\end{equation}
where $p(d) \equiv p(d_1...d_N)$, provided that we expect our model parameters $\vec{\theta}$ to be identical for all N events. If this is not the case \emdashb if each event is expected to be described by its own individual $\vec{\theta_i}$, where each $\vec{\theta_i}$ is informed by a shared process or characteristic across all events \emdashb then we must turn to a more general case of \textit{hierarchical} Bayesian inference.

In hierarchical Bayesian inference, each event's individual model parameters $\vec{\theta}_i$ are drawn from some common distribution described by a shared parameter (or parameters) $\phi$. This creates a hierarchy where the population-level parameter $\phi$ informs the individual event parameters $\vec{\theta}_i$, which in turn describe the data $d$. Using Bayes' theorem and the definition of conditional probability $p(\vec{\theta},\phi) = p(\vec{\theta}|\phi)p(\phi)$, we can write for a single event
\begin{equation}
\begin{split}
    p(\vec\theta,\phi|d) & =  \frac{p(d|\vec\theta,\phi)p(\vec\theta,\phi)}{p(d)}\\
    & = \frac{p(d|\vec\theta)p(\vec\theta|\phi)p(\phi)}{p(d)}\\
    & \propto p(d|\vec\theta)p(\vec\theta|\phi)p(\phi).
\end{split}
\end{equation}
This hierarchy is often described in shorthand using the following notation:
\begin{equation}
\begin{split}
    d | \vec\theta & \sim p(d|\vec\theta),\\
    \vec\theta | \phi & \sim p(\vec\theta|\phi),\\
    \phi & \sim p(\phi).
\end{split}
\end{equation}
To generalize to $N$ events, we can follow the same procedure as we did for non-hierarchical inference, taking a product of the individual event likelihoods:
\begin{equation}
\begin{split}
    p(\vec\theta_{1...N},\phi|d_{1...N}) & = \frac{1}{p(d)} \prod_{i=1}^N \mathrlap{\left[ p(d_i|\vec{\theta}_i)p(\vec\theta_i|\phi) \right]p(\phi)}\\
    & \propto \prod_{i=1}^N \left[p(d_i|\vec{\theta}_i)p(\vec\theta_i|\phi) \right]p(\phi).
\end{split}
\end{equation}
where again $p(d) \equiv p(d_1...d_N)$. This returns a population-informed posterior on both the individual event parameters $\vec\theta_i$ and the population-level parameter, $\phi$. If we are ultimately only interested in  $\phi$, we can marginalize over $\vec\theta_i$, treating them as ``nuisance", or latent parameters and recovering a marginalized posterior distribution on $\phi$ only:
\begin{equation}
    p(\phi|d_{1...N}) \propto \prod_{i=1}^N \left[\int p(d_i|\vec\theta_i)p(\vec\theta_i|\phi)d\vec\theta_i \right]p(\phi).
\end{equation}\label{eq:generic_HB_final}

Hierarchical Bayesian methods are often applied in astronomy to situations where a set of astrophysical objects or events have differing individual properties which all stem from the same underlying physical processes. Examples include the spatial distribution of trans-Neptunian objects \citep{loredo_accounting_2004}, exoplanet populations with a shared mass-radius relationship \citep{wolfgang_probabilistic_2016}, galactic satellites whose orbital dynamics are dominated by their host galaxy \citep{martinez_robust_2015}, ensembles of BNS inspiral signals with tidal signatures determined by $\Lambda(m)$ \citep{hernandez_vivanco_measuring_2019}, observations of compact binary mergers arising from an underlying population distribution and cosmology \citep{mandel_extracting_2019,abbott_constraints_2021,abbott_population_2022} or, more pertinently, BNS post-merger remnants with various individual values of $\mc$ and $\fpeak$ but a shared EoS.\\

In this work, we present a method for combining post-merger information from an ensemble of BNS mergers through a hierarchical Bayesian schema, in order to constrain $\R$. We utilize quasi-universal, EoS-agnostic empirical relations based on those of \citet{vretinaris_empirical_2020} to connect posterior distributions for $\mc$ and $\fpeak$ to the EoS proxy $\R$, allowing for collective inference of the latter quantity across the ensemble of BNS events. This approach is presented in full in \S\ref{methods} and its efficacy is explored in \S\ref{results}, including application to simulations of the next two LVK observing runs (O4 and O5) and an extended run at LIGO A+ sensitivity.

\section{Methods}\label{methods}

\subsection{Empirical Relations}\label{empiricalrelations}
\begin{figure}
    \centering
    \includegraphics[width=1\linewidth]{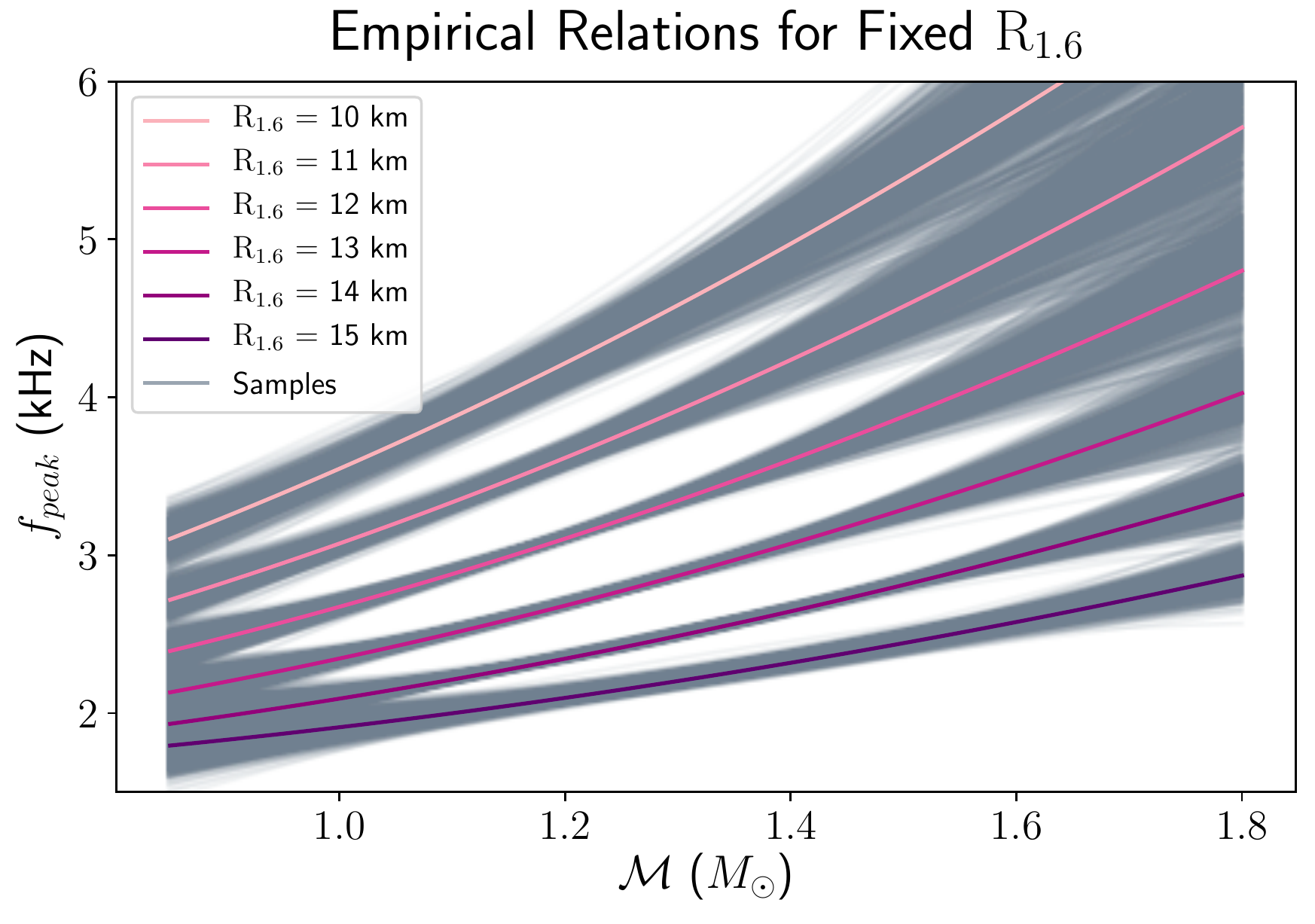}
    \caption{Distribution of $N_{\mathrm{R}}=1000$ sampled relations, evaluated at a range of values of $\R$, alongside our fitted empirical relations, likewise evaluated over the same range. The sampled distribution is consistent with the fit relation, and characterizes the relation's intrinsic uncertainty.}
    \label{fig:ER_dist}
\end{figure}

It has been shown via large numbers of numerical relativity simulations that there are quasi-universal, EoS-agnostic empirical relations connecting $\fpeak$, $\mc$, and $\R$ \citep{bauswein_measuring_2012,bauswein_equation--state_2012,vretinaris_empirical_2020}. While \citet{vretinaris_empirical_2020} have calculated a wide variety of empirical relations for $R_{1.\text{x}}\; (\text{x} \in [2,4,6,8])$ and several different post-merger signal characteristics, we use a relation of the form $\fpeak = f(\R,\mc)$, based on Equation 4 of \citet{vretinaris_empirical_2020},\footnote{Recently, it has been speculated that these relations may be less tight, especially for EoSs with significant positive M-R slope \citep{raithel_characterizing_2022}. However, a re-analysis of previous results does not support this conclusion for microphysical models with similar variations of the slope in the M-R curves \citep{bauswein_equation--state_2012}. Also, using the large sample of microphysical EoSs of \citet{vretinaris_empirical_2020}, we do not find evidence that the introduction of an $R_{1.4}/R_{1.8}$ term as proposed in \citet{raithel_characterizing_2022} would improve the fits (for an optimally chosen radius) nor do we find the same magnitude of scatter in the relations for $\fpeak$ (see also \citet{lioutas_frequency_2021} for a discussion of frequency deviations).} which possessed the lowest residuals of the relations considered in \citet{vretinaris_empirical_2020}. However, any of the relations for $R_{1.\text{x}}$ could be used with this procedure. The original relation of \citet{vretinaris_empirical_2020} specifies a model of the chirp-mass-scaled peak frequency $\fpeak/\mc$ as a function of $\mc$ and $\R$; that is, $\fpeak/\mc=f(\mc,\R)+\epsilon$, where $\epsilon\sim \mathcal{N}(0,\sigma^2)$ is the error term. This exact form would suffice if we only needed a point estimate of $\fpeak$ as a function of $\mc$ and $\R$, as we could simply multiply both sides by $\mc$. However, we must account for the uncertainty in the fitted values we obtain from this relation, and multiplying both sides by $\mc$ produces a relation of the form $\fpeak = g(\mc,\R) + \mc\times\epsilon$, which has an error term dependent on $\mc$. To avoid this, we consider a modified relation, based on the original \citet{vretinaris_empirical_2020} relations, of the form
\begin{equation}\label{eqn:empirical_relation}
    \begin{split}
        \fpeak = \beta_0\  &+ \beta_1\mc + \beta_2\mc^2 + \beta_3\R\mc\\
    &+ \beta_4\R\mc^2 + \beta_5\R^2\mc, 
    \end{split}
\end{equation}
with an additional error term $\epsilon\sim \mathcal{N}(0,\sigma^2)$, where $\beta_0$ through $\beta_5$ are our fit coefficients, and $\fpeak$, $\mc$, and $\R$ are in kHz, $\msun$, and km, respectively. Fitting this model to the data in Tables I and II in \citet{vretinaris_empirical_2020} using least squares, we obtain the fit coefficients in Table \ref{tbl:coeff} with a maximum residual of $0.185$ kHz and a residual standard error $0.061$ kHz. We obtain a set of relations that is consistent with those of \citet{vretinaris_empirical_2020} to within the intrinsic error of their relations. We also obtain the covariance matrix $\hat\Sigma$ for the vector of estimated coefficients $\hat\beta=(\hat\beta_0,\ldots,\hat\beta_5)$. This enables us to characterize the intrinsic empirical relation uncertainty by sampling $N_{\mathrm{R}}=1000$ relations from the multivariate normal distribution $\mathcal{N}(\hat\beta,\hat\Sigma)$. These fits and their corresponding sampled relations are shown in Fig. \ref{fig:ER_dist}.

\begin{table}\label{tbl:coeff}
    \centering
    \begin{tabular}{c D{.}{.}{1,4} l}%l D{.}{.}{1,4}}
        \hline
        \multicolumn{1}{c}{Coefficient} & \multicolumn{1}{c}{Value} &
        \multicolumn{1}{l}{Unit}\\
        \hline
        $\beta_0$ &  1.5220 & kHz\\
        $\beta_1$ &  8.4021 & kHz/$\msun$ \\
        $\beta_2$ &  2.3876 & kHz/$\mathrm{M}_{\odot}^2$\\
        $\beta_3$ & -1.1133 & kHz/(km $\msun$)\\
        $\beta_4$ & -0.1291 & kHz/(km $\mathrm{M}_{\odot}^2$)\\
        $\beta_5$ &  0.0366 & kHz/($\mathrm{km}^2$ $\msun$)\\
        \hline
    \end{tabular}
    \caption{Fit coefficients for the relation of Eq.~\eqref{eqn:empirical_relation}.}
\end{table}

It is worth noting that these relations are derived for $\mc$ and $\fpeak$ in the source rest frame. These quantities as measured by GW detectors include some amount of cosmological redshift, which must be accounted for in order to accurately employ the empirical relations. Inspiral source frame $\mc$ posteriors are a standard data product, but we need to manually correct for redshift in our $\fpeak$ posteriors. This correction is of order 5-10\% in $\fpeak$ for the most distant (and therefore least informative) BNS events we consider in this study. For the purposes of this work, we make the simplifying assumption that the redshift of each event is known. However, we note that the extension of our formalism to include inspiral redshift posteriors is straightforward and should be incorporated in the future.\footnote{Not only will next-generation detectors be able to observe more events at appreciable redshifts, but \citet{haster_inference_2020} also showed that high-redshift post-merger signals contribute more information due to the peak frequency being redshifted towards the detectors' more sensitive band.}

% I hope whoever runs @LeaksPh has a nice day

\subsection{Derivation of Hierarchical Framework}\label{formalism}
We wish to glean whatever information on $\R$ is contained in the GW observations of an ensemble of BNS mergers. We can do this by considering measurements of the chirp mass (via the inspiral signal; see \S \ref{mchirp-approx}) and the post-merger peak frequency (see \S \ref{bayeswave}), connected through the empirical relations of \citet{vretinaris_empirical_2020} (see \S\ref{empiricalrelations}). We construct a statistical hierarchy such that
\begin{equation}
\begin{split}
    D_i | \mci, \fpi & \sim p(D_{\mathrm{IN},i}|\mci)p(D_{\mathrm{PM},i}|\fpi),\\
    \mci,\fpi|\R & \sim p(\mci,\fpi|\R),\\
    \R & \sim \pi(\R),\\
\end{split}
\end{equation}
where $D_i$ is the collected GW data of the $i^{th}$ BNS merger, consisting of its inspiral data $D_{\mathrm{IN},i}$ and post-merger data $D_{\mathrm{PM},i}$. The first line incorporates the inspiral and post-merger analyses, and makes an important assumption: that the chirp mass as measured from the inspiral signal is conditionally independent of the post-merger peak frequency from the post-merger signal. Despite the fact that both measured quantities arise from the same BNS merger, this is a reasonable assumption. While $\mc$ and $\fpeak$ are related physically, the data analysis approaches leading to their posterior distributions are independent of each other (see \S \ref{bayeswave} and \S \ref{mchirp-approx}). Parameter estimation performed on the inspiral signal does not account for post-merger GWs, and is not biased by the presence/absence or morphology of a post-merger signal \citep{dudi_relevance_2018}. Similarly, the morphology-independent analysis used to reconstruct the post-merger signal and compute the peak frequency posterior is entirely agnostic of all system parameters, including the chirp mass. It would be inaccurate to say that $\mc$ and $\fpeak$ are unrelated, but their measurements can be approximated as statistically independent in the low-SNR limit.\footnote{For louder signals, this issue can be handled by simultaneous analysis of the inspiral and post-merger signal \citep{wijngaarden_probing_2022}.}

We additionally neglect selection effects arising from the increased sensitivity of our detectors to BNS mergers with higher values of $\mc$. In practice, this effect will need to be accounted for to avoid bias; however, the means of doing so is known \citep{mandel_extracting_2019} and will be incorporated in the future.

Ultimately, we want to calculate the posterior for $R_{1.6}$ given $N$ events' post-merger and inspiral data ($D_{\mathrm{PM}}$ and $D_{\mathrm{IN}}$, respectively):
\begin{equation}
    p(\R|D) = p(\R|(D_{\mathrm{PM}},D_{\mathrm{IN}})_{i...N}).
\end{equation}
This is expanded into a likelihood and a prior:
\begin{equation}
    p(\R|D) =  \frac{p((D_{\mathrm{PM}},D_{\mathrm{IN}})_{i...N}|\R)}{p(D)}\pi(\R).
\end{equation}
The likelihood for $N$ events can be written as
\begin{widetext}
\begin{equation}\label{like_with_marg}
\begin{split}
        p(D|\R) & = \prod_{i=1}^N p((D_{\mathrm{PM}},D_{\mathrm{IN}})_i|\R)\\
         & = \prod_{i=1}^N \bigg[\int \int p(D_{\mathrm{IN},i},D_{\mathrm{PM},i}|\mci,\fpi) p(\mci,\fpi|\R)d\fpi d\mci \bigg]\\
         & = \prod_{i=1}^N \bigg[\int \int p(D_{\mathrm{IN},i}|\mci)p(D_{\mathrm{PM},i}|\fpi) p(\mci,\fpi|\R)d\fpi d\mci \bigg],
\end{split}
\end{equation}
\end{widetext}
where we have introduced a marginalization over $\fpeak$ and $\mc$ (as we are only interested in $\R$) and incorporated the assumption that the inspiral and post-merger measurements are conditionally independent. As the distributions available on $\fpeak$ and $\mc$ are \textit{posteriors} arising from established GW analyses (see \S\ref{bayeswave}, \S\ref{mchirp-approx}), we use Bayes' theorem to expand each of our parameter likelihoods $p(D_{\mathrm{PM},i}|\fpi)$ and $p(D_{\mathrm{IN},i}|\mci)$ as follows:
\begin{widetext}
\begin{equation}\label{eqn:posterior-inversion}
p(D|\R) = \prod_{i=1}^N \bigg[\int \int \frac{p(\mci|D_{\mathrm{IN},i})p(D_{\mathrm{IN},i})}{p_0(\mci)}\frac{p(\fpi|D_{\mathrm{PM},i})p(D_{\mathrm{PM},i})}{p(\fpi)}p(\fpi,\mci|\R)d\fpi d\mci \bigg].
\end{equation}
\end{widetext}
To expand the likelihoods for $\fpeak$ and $\mc$ in this way, the $p_0(\mc)$ and $p(\fpeak)$ priors we use must be identical to those used in the initial parameter estimation process. Finally, we expand $p(\fpi,\mci|\R)=p(\fpi|\mci,\R)p(\mci)$, assuming that the distribution of $\mc$ is independent of $\R$; that is, $p(\mc|\R) = p(\mc)$.\footnote{The astrophysical distribution of NS masses is currently uncertain~\citep{landry_mass_2021,abbott_population_2022}, including whether the maximum NS mass is determined by the EoS through $M_{\mathrm{TOV}}$ (the Tolman–Oppenheimer–Volkoff limit for the maximum mass of a non-rotating NS) or astrophysical processes. At current statistical uncertainties these assumptions have a small impact on the inferred $M_{\mathrm{TOV}}$; however, they are negligible for the radius~\citep{legred_impact_2021}. Though the impact on $M_{\mathrm{TOV}}$ will become more prominent with additional data, it has been shown that \emdashb for the pertinent case of generic, flexible analyses \emdashb $\R$ is largely independent of both $M_{\mathrm{TOV}}$ and such astrophysical considerations~\citep{legred_implicit_2022}, allowing for the assertion that $p(\mc|\R) = p(\mc)$.} In practice, one should parameterize $p(\mc)$ and simultaneously infer $\R$ alongside the astrophysical $\mc$ distribution; however, we assume for simplicity that $p(\mc)$ is known. At this juncture we must note the distinction between $p_0(\mc)$, the prior for $\mc$ used in the original inspiral analyses from which we derive our $\mc$ posteriors, and $p(\mc)$, our own choice of prior for $\mc$. While $p_0(\mc)$ and $p(\mc)$ can be identical, in a realistic case $p_0(\mc)$ will have been chosen by others, and may or may not align with our own choice of prior.\footnote{In the realistic case of a future post-observing-run analysis, we would use the published $\mc$ posterior distributions for each BNS event. This would then necessitate that we use the $\mc$ prior $p_0(\mc)$ chosen for these analyses \emdashb likely the standard uniform priors used in previous LVK catalogues (see, e.g., GWTC-3 \cite{abbott_gwtc-3:_2021}) \emdashb to ensure that we correctly recover the likelihood from the posterior samples available to us. However, our choice of $p(\mc)$is not constrained in this fashion and should instead describe our expectations of the overall population distribution of $\mc$.} The specific distributions for $p_0(\mc)$ and $p(\mc)$ used in this work are discussed in \S\ref{priorchoice}.

We can now write the posterior for $\R$, factoring out the individual event evidences in favor of the usual proportionality:
\begin{widetext}
\begin{equation}\label{initial_posterior}
\begin{split}
    p(\R|D) &= \pi(\R)\frac{\prod_{k=1}^N \big[p(D_{\mathrm{IN},k})p(D_{\mathrm{PM},k})\big]}{p(D)}\\
    &\quad\times\prod_{i=1}^N \bigg[\int \int  \frac{p(\mci|D_{\mathrm{IN},i})}{p_0(\mci)} \frac{p(\fpi|D_{\mathrm{PM},i})}{p(\fpi)} p(\fpi|\mci,\R)p(\mci)d\fpi d\mci \bigg]\\
     &\propto \pi(\R)\prod_{i=1}^N \bigg[\int \int  \frac{p(\mci|D_{\mathrm{IN},i})}{p_0(\mci)} \frac{p(\fpi|D_{\mathrm{PM},i})}{p(\fpi)} p(\fpi|\mci,\R)p(\mci)d\fpi d\mci \bigg].
\end{split}
\end{equation}
\end{widetext}

The term $p(\fpi|\mci,\R)$ serves to tie our $\fpeak$ and $\mc$ posteriors to $\R$ while accounting for the intrinsic scatter of the empirical relations. As discussed in \S\ref{empiricalrelations}, we model the uncertainty of the empirical relation by assuming that the coefficients in Eq.~\eqref{eqn:empirical_relation} are well modeled by a multivariate normal distribution $\mathcal{N}(\hat\beta,\hat\Sigma)$ and sampling from this distribution. Formally, let
\begin{equation}
    p(\fpi|\mci,\R)=\int p(\fpi|\mci,\R, F_E)p(F_E)dF_E,
\end{equation}
where the integral is a marginalization over the intrinsic uncertainty of the empirical relations, represented by a distribution of potential empirical relations $F_E$. We approximate this using our empirical relation samples $F_{E,j}$ such that
\begin{equation}
    p(\fpi|\mci,\R)=\frac{1}{N_{\mathrm{R}}} \sum_j p(\fpi|\mci,\R, F_{E,j}),
\end{equation}
where $N_{\mathrm{R}}$ is the number of empirical relation samples.
We take an empirical Bayes approach and assume that the prediction of each sampled relation is exact; that is, 
\begin{equation}
    p(\fpi|\mci,\R,F_{E,j})=\delta(\fpi - \hfpij), 
\end{equation}
where the $\delta$ function is a Dirac delta and $\hfpij = F_{E,j}(\mci,\R)\times(1+z)^{-1}$, the $j^{th}$ sampled relation's redshift-corrected\footnote{We assume the redshift is known; see \S\ref{empiricalrelations}.} prediction for $\fpi$ as a function of $\mc$ for a given value of $\R$. The posterior then becomes

\begin{widetext}
\begin{equation}
\begin{split}
    p(\R|D) \propto &\  \pi(\R) \prod_{i=1}^N \bigg[\int\int \frac{p(\mci|D_{\mathrm{IN},i})}{p_0(\mci)} \frac{p(\fpi|D_{\mathrm{PM},i})}{p(\fpi)} \frac{1}{N_{\mathrm{R}}}\sum_j\big[\delta(\fpi - \hfpij)\big] p(\mci)d\fpi d\mci \bigg]\\
    \propto &\  \pi(\R) \prod_{i=1}^N \bigg[\int \frac{1}{N_{\mathrm{R}}}\sum_j \bigg(  \frac{p(\mci|D_{\mathrm{IN},i})}{p_0(\mci)} \frac{p(\hfpij|D_{\mathrm{PM},i})}{p(\hfpij)}\bigg)p(\mci) d\mci \bigg].\\
\end{split}
\end{equation}
\end{widetext}

Finally, we must account for an additional subtlety: we only consider data on a [1.5,~4]~kHz frequency band. This choice is partially driven by the fact that, at frequencies above 4 kHz, the rapid degradation of the detector sensitivity becomes a severely limiting factor. More importantly, all numerical relativity simulations used to fit the \citet{vretinaris_empirical_2020} relations have peak frequencies within this band; relying on the empirical relations' predictions outside of this regime would necessarily comprise extrapolation. Thankfully, given the broad coverage of the \citet{vretinaris_empirical_2020} data, we can be assured that few (if any) systems will fall outside of this frequency band. 

Despite this choice, some combinations of $\R$ and $\mc$ \emdashb especially for $\R\lesssim10$ km \emdashb yield an empirical-relation-predicted $\hfp$ that lies outside this range. As we do not consider data from these out-of-band frequencies, any $(\mci,\R)$ pair that results in such a prediction should  not encode any additional probability, instead returning only the prior probability on $\R$.\footnote{This can be considered an ``analysis selection effect'' (as opposed to the more typical ``observational selection effect'') wherein some data is unanalyzed/unobserved depending on what the \emph{observed} (rather than the true) parameters are.} Accordingly, we split the marginalization over $\mc$ into two integrals: one for each of the in-band ($\mathcal{I}$) and out-of-band ($\mathcal{O}$) cases. We can then write

\begin{widetext}
\begin{equation}
\begin{split}
    p(\R|D) \propto \pi(\R) \prod_{i=1}^N & \bigg[\int_{\mathcal{I}} \frac{1}{N_{\mathrm{R}}} \sum_j\bigg( \frac{p(\mci|D_{\mathrm{IN},i})}{p_0(\mci)} \frac{p(\hfpij|D_{\mathrm{PM},i})}{p(\hfpij)} \bigg)p(\mci) d\mci\\
    &\qquad\qquad\qquad\qquad\qquad\ +\  \int_{\mathcal{O}}\frac{1}{N_{\mathrm{R}}}\sum_j\bigg(\frac{p(\mci|D_{\mathrm{IN},i})}{p_0(\mci)} \frac{p(\hfpij|D_{\mathrm{PM},i})}{p(\hfpij)} \bigg) p(\mci)d\mci \bigg],\\
\end{split}
\end{equation}
\end{widetext}

where the integrals over $\mathcal{I}$ and $\mathcal{O}$ integrate over \textit{only} $\mc$ values associated with in-band and out-of-band values of $\hfpij(\mci,\R)$, respectively. Finally, we note that if $\hfpij$ is out-of-band, $p(\hfpij|D_{\mathrm{PM},i})$ is independent of the post-merger signal. That is, $p(\hfpij|D_{\mathrm{PM},i})=p(\hfpij)$, so \\
\begin{equation}\label{eqn:fpeak-uninformative}
    \sum_j\frac{p(\hfpij|D_{\mathrm{PM},i})}{p(\hfpij)} = \sum_j\frac{p(\hfpij)}{p(\hfpij)} = N_{\mathrm{R}} \times 1.
\end{equation}
This then (at long last!) allows us to write the final form of our posterior:
\begin{widetext}
\begin{equation}\label{eqn:finalpost_propto}
    p(\R|D) \propto \pi(\R) \prod_{i=1}^N \bigg[ \int_{\mathcal{I}} \frac{1}{N_{\mathrm{R}}}\sum_j\bigg(\frac{p(\mci|D_{\mathrm{IN},i})}{p_0(\mci)} \frac{p(\hfpij|D_{\mathrm{PM},i})}{p(\hfpij)}\bigg) p(\mci)d\mci + \int_{\mathcal{O}}\bigg( \frac{p(\mci|D_{\mathrm{IN},i})}{p_0(\mci)}\bigg) p(\mci)d\mci \bigg].
\end{equation}
\end{widetext}

Each component of Eq.~\eqref{eqn:finalpost_propto} is explored in detail below. The process of calculating the peak frequency prior and posterior via BayesWave \citep{cornish_bayeswave_2015} is described is \S \ref{bayeswave}; the approximation we employ to simulate the chirp mass posteriors from inspiral analyses is discussed in \S\ref{mchirp-approx}; the choice of priors is considered in \S \ref{priorchoice}. Finally, the implementation of this formalism in the \textsc{bayestack} package is detailed in \S\ref{code}.

\subsection{BayesWave: Recovering the Post-merger Peak Frequency}\label{bayeswave}

\begin{figure}
    \centering
    \includegraphics[width=0.9\linewidth]{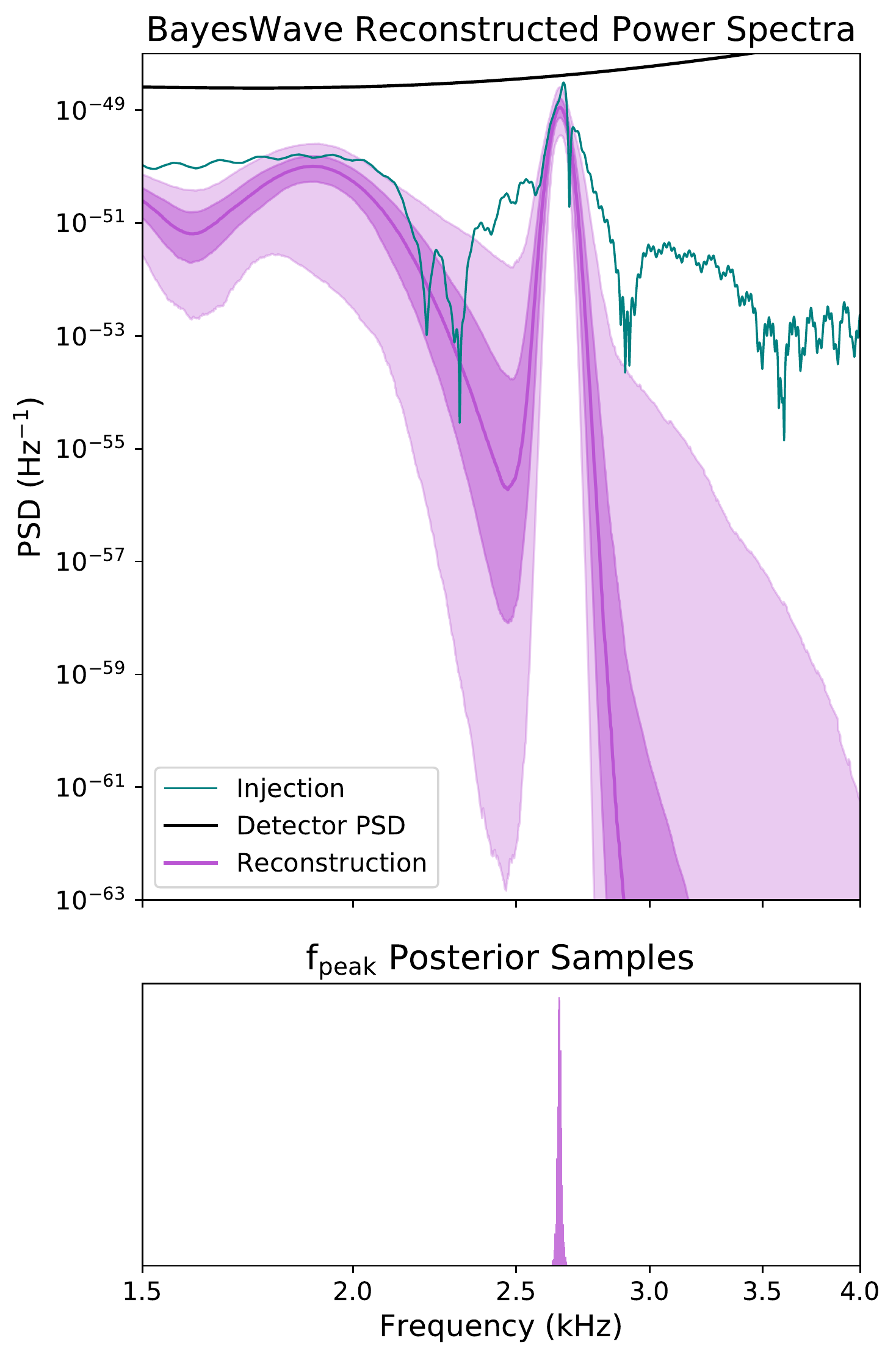}
    \caption{BayesWave signal reconstruction and corresponding peak frequency posterior for a strong BNS post-merger signal. This BayesWave run was performed on a SFHX post-merger waveform with $m_1=1.03 \msun$ and $m_2=1.09 \msun$ injected at 30~Mpc at NEMO sensitivity. In the top panel, the injected waveform is in teal and the median BayesWave reconstruction with 50\% (dark shading) and 90\% (light shading) C.I. is shown in purple. The post-merger signal is reconstructed nicely, and the injected waveform's peak frequency $\fpeak=2.67$ kHz is recovered to within $\sim0.01$ kHz.}
    \label{fig:BW-strong}
\end{figure}

\begin{figure}
    \centering
    \includegraphics[width=0.9\linewidth]{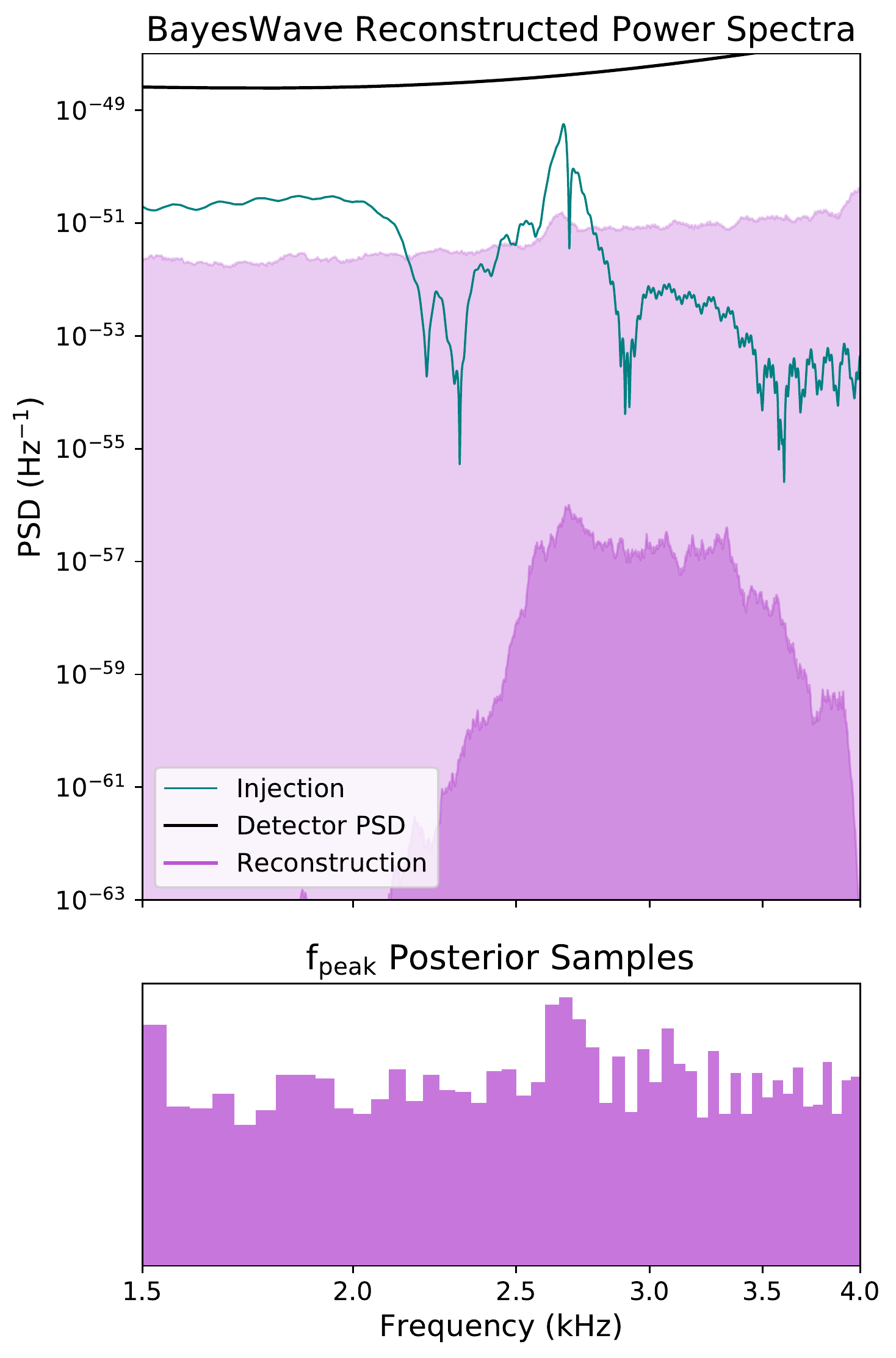}
    \caption{BayesWave signal reconstruction and corresponding peak frequency posterior for a marginal BNS post-merger signal. This BayesWave run was performed on a post-merger waveform with $m_1=1.03 \msun$ and $m_2=1.09 \msun$ injected at 70~Mpc at NEMO sensitivity. In the top panel, the injected waveform is in teal and the 50\% (dark shading) and 90\% (light shading) C.I. of the BayesWave reconstruction are shown in purple. No median reconstruction is shown because the reconstructed signal is consistent with zero. Despite this, the spectrum and corresponding $\fpeak$ posterior show a slight preference for the injected waveform's peak frequency of 2.67 kHz. Taken by itself, this recovery of the peak frequency is unconvincing \emdashb and would be considered a nondetection \emdashb but nonetheless contains information about the post-merger signal.}
    \label{fig:BW-weak}
\end{figure}

To obtain posterior distributions for the post-merger peak frequency, we use the  BayesWave package \citep{cornish_bayeswave_2015,littenberg_bayesline_2015,cornish_bayeswave_2021} and follow the methods of \citet{chatziioannou_inferring_2017} and \citet{wijngaarden_probing_2022}. 
 In the BayesWave algorithm, each potential signal is constructed from an arbitrary number of Morlet-Gabor wavelets with associated wavelet parameters, accounting for sky location, polarization angle, and ellipticity. BayesWave uses a reversible jump Markov chain Monte Carlo algorithm \citep{green_reversible_1995} to sample the joint and variable-dimension posterior of these parameters. The resulting distribution of signal reconstructions maps to a posterior distribution of peak frequencies, where each $\fpeak$ sample is the spectrum maximum on [1.5,~4]~kHz of its corresponding signal reconstruction sample\footnote{Power contribution from the inspiral and merger is suppressed via windowing in time-domain; see \citet{chatziioannou_inferring_2017} for further detail.} \citep{chatziioannou_inferring_2017}. If a given signal reconstruction sample has no such maximum, a peak frequency sample is instead drawn from the $\fpeak$ prior. Sample BayesWave waveform reconstructions and their corresponding peak frequency posteriors are shown in Figs. \ref{fig:BW-strong} and \ref{fig:BW-weak} for strong and weak post-merger signals, respectively, using the sensitivity of the proposed Neutron Star Extreme Matter Observatory (NEMO) detector.

We model each BayesWave peak frequency posterior by fitting the posterior samples with a \textsc{PESummary} \citep{hoy_pesummarypesummary_2021} reflection-bounded kernel density estimator (KDE). The use of KDEs to model the underlying probability distribution associated with a set of marginal posterior samples introduces an inherent approximation, but allows for rapid post-processing of data products from otherwise time-intensive parameter estimation processes. Accordingly, this approximation is often employed in hierarchical gravitational-wave data analyses (e.g., \citet{lackey_reconstructing_2015,golomb_hierarchical_2022}), although other approximate representations have been proposed, including interpolation of the individual marginal likelihoods with random forest regressors \citep{hernandez_vivanco_measuring_2019} or Gaussian processes \citep{lange_rapid_2018}. Employing a reflection-bounded KDE allows us to mitigate the edge effects otherwise characteristic of standard KDEs at the edges of bounded distributions such as those of our $\fpeak$ posteriors.

\subsection{Inspiral Chirp Mass Posteriors}\label{mchirp-approx}

We employ a convenient approximation for the chirp mass posteriors of each event obtained from the inspiral signal. As shown in \citet{farr_parameter_2016}, the relative width of the chirp mass posterior distribution scales inversely with SNR. That is, $\sigma_{\mc}/\mc \propto (\textrm{SNR})^{-1}$. We therefore approximate our chirp mass posteriors as draws from $\mathcal{N}(\mu_{\mc},\sigma_{\mc})$, where $\mu_{\mc}$ is itself drawn from $\mathcal{N}(\mcit,\sigma_{\mc})$ to represent the scatter in the $\mc$ posterior mean due to detector noise and $\sigma_{\mc}$ is defined using the chirp mass posterior width $\sigma_{\textrm{170817}}=0.004\msun$,\footnote{This value of $\sigma_{\textrm{170817}}=0.004\msun$ is the upper 90\% C.I. bound; we are being conservative in our estimate of the chirp mass precision.} chirp mass $\mc_{\textrm{170817}}=1.188\msun$, and network SNR $\textrm{SNR}_{170817} = 32.4$ of the BNS merger GW170817 as our point of reference \citep{abbott_gw170817_2017} such that
\begin{equation}\label{eqn:chirp-mass-scaling}
    \sigma_{\mci} = \frac{\mci}{\mc_{\textrm{170817}}}\frac{\textrm{SNR}_{170817}}{\textrm{SNR}_i}\sigma_{\textrm{170817}}.
\end{equation}
We simulate $N_{\mathrm{R}}=1000$ samples from the resulting distribution for each event and fit a KDE to the samples as discussed in \S\ref{bayeswave}. Simulating our chirp mass posteriors in this way allows us to maintain the expedience afforded by such an approximation while retaining the ability to generalize to real data products in the future.\\ 

\subsection{Choice of Priors}\label{priorchoice}
\begin{figure}
    \centering
    \includegraphics[width=0.9\linewidth]{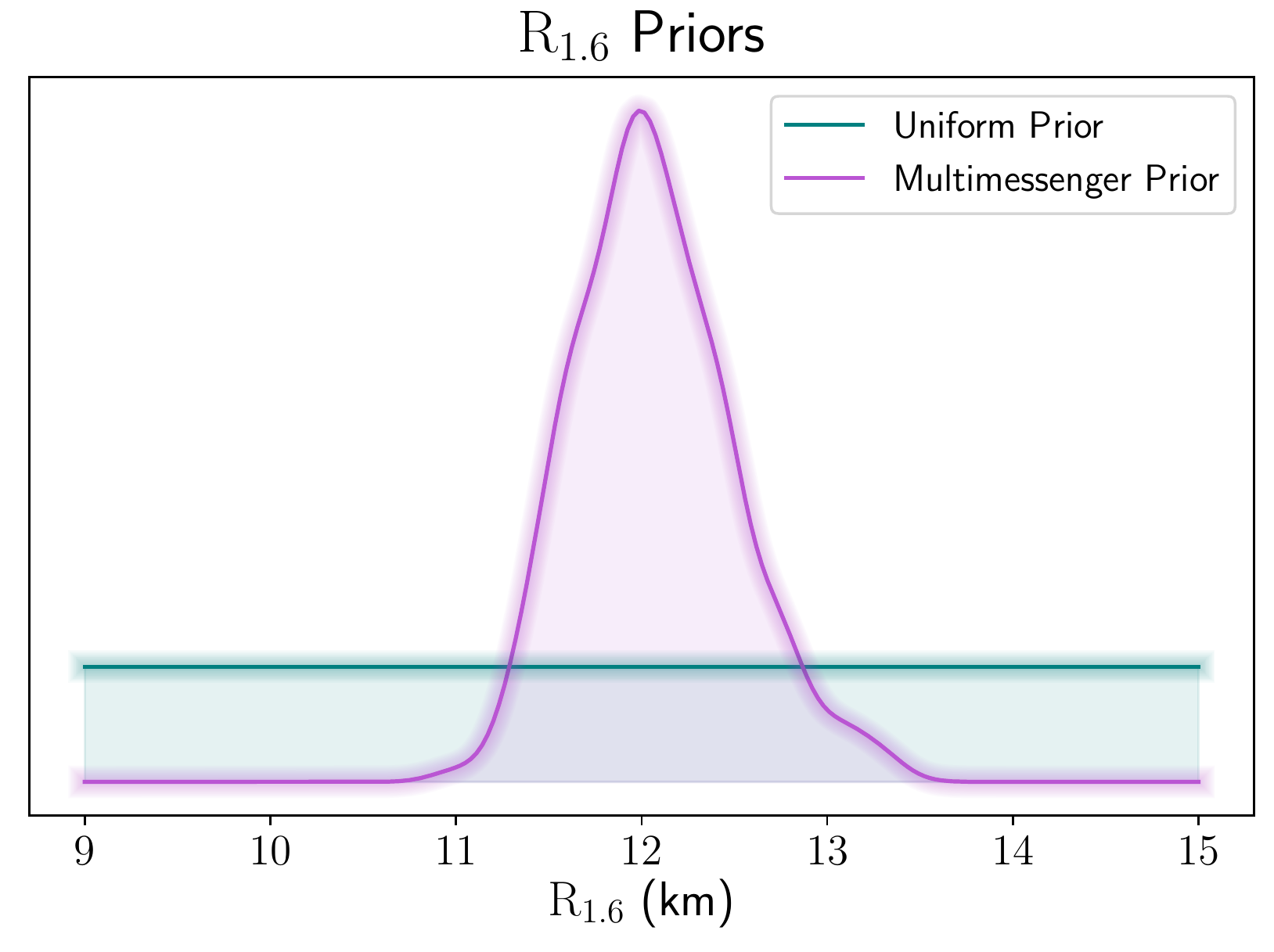}
    \caption{$\R$ priors used in this work. One is a uniform prior on $[9,15]$ km, the other is an astrophysically motivated $\R$ prior, derived from the multimessenger analysis of \citet{huth_constraining_2022}.}
    \label{fig:prior-dist}
\end{figure}

For $\R$, we consider both uniform and astrophysically motivated priors. The former will allow us to consider our results in isolation, whereas the latter will demonstrate what we may be able to contribute within a larger multimessenger framework. The uniform prior is on $[9,15]$ km. The astrophysical prior uses the posterior distribution from the multimessenger constraints on $\R$ found in \citet{huth_constraining_2022}. This distribution is consistent with current constraints on the NS EoS and was derived through a combination of maximum NS mass constraints from observed pulsars and the remnant classification of GW170817, NICER NS mass measurements, GW analyses of the BNS mergers GW170817 and GW190425 along with their electromagnetic (EM) counterpart (AT2017gfo in the case of GW170817) or lack thereof (in the case of GW190425), and data from terrestrial heavy-ion collision experiments \citep{huth_constraining_2022}. While the results presented in \citet{huth_constraining_2022} are for $\text{R}_{1.4}$ and $\text{R}_{2.0}$ only, the authors provided all considered EoSs and their associated posterior probabilities, from which any $\text{R}_{1.\textrm{x}}$ can be calculated, and from which we obtain a distribution on $\R$ specifically. Both priors are shown in Fig. \ref{fig:prior-dist}.

\begin{figure}
    \centering
    \includegraphics[width=0.9\linewidth]{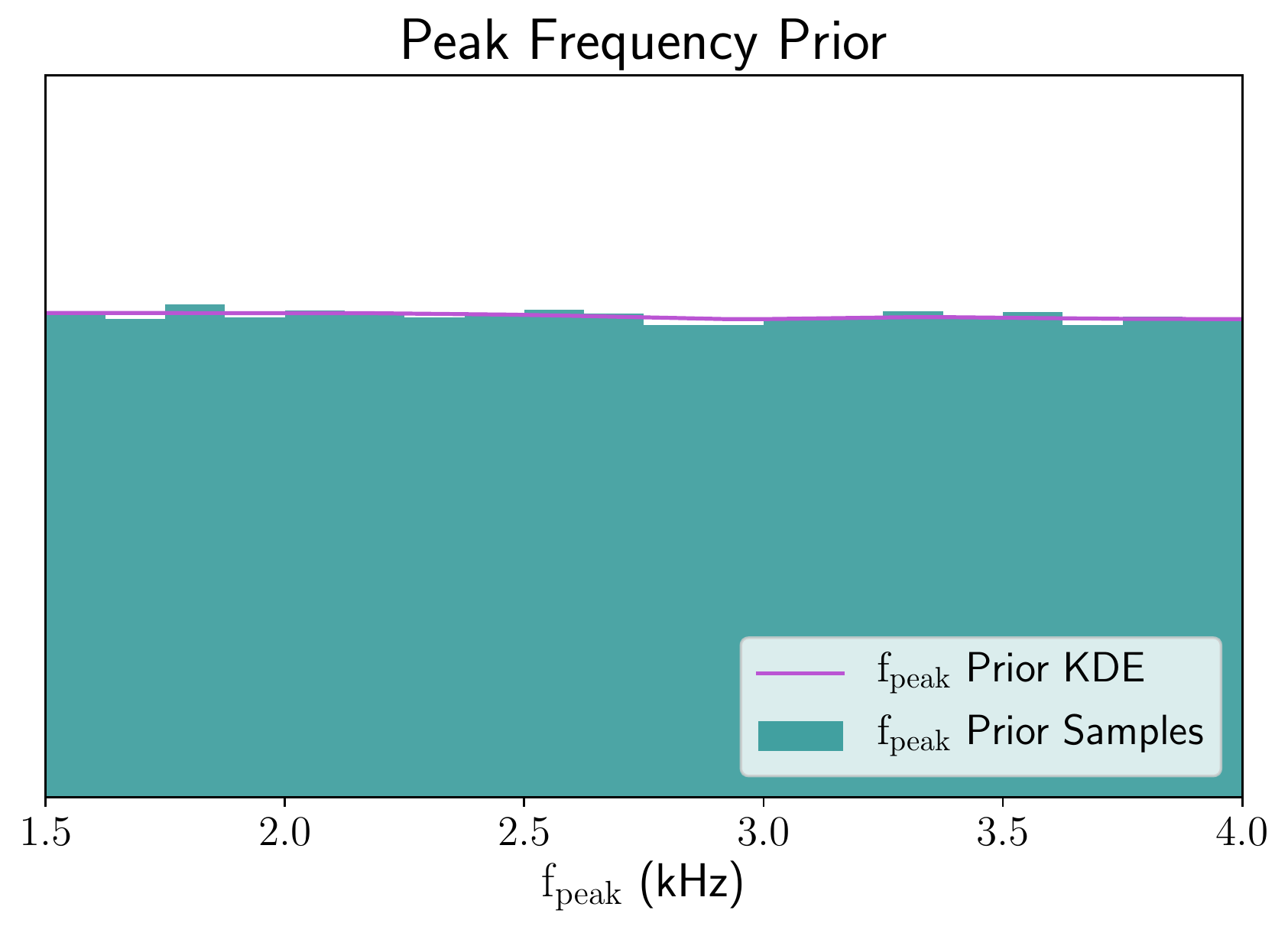}
    \caption{Peak frequency prior samples, with a reflection-bounded KDE approximation of its probability distribution. These samples were computed from an ensemble of N=50 BayesWave analyses over simulated instrumental noise.}
    \label{fig:fpeak-prior}
\end{figure}

Eq.~\eqref{eqn:finalpost_propto} further includes two priors on $\mc$ \emdashb the original analysis prior $p_0(\mc)$ and our own prior $p(\mc)$ \emdashb and one on $\fpeak$. We are free to choose $p(\mc)$, and let $p(\mc) = \mathcal{N}(\mu=1.33\msun,\sigma=0.09\msun)$ to match the distributions used in \citet{abbott_prospects_2020-1} and \citet{petrov_data-driven_2022} as we use the latter to simulate our event catalogues in \S\ref{obs_run_sims}. We note that the choice of $p(\mc)$ has minimal impact on the final results of our analysis due to the precision of inspiral $\mc$ posteriors. However, we emphasize that $p_0(\mc)$ and $p(\fpeak)$ are intrinsically tied to the choices made in the original analyses for $\mc$ and $\fpeak$, respectively, solely serving to cancel out the original priors used to produce our posterior samples. Determining the original $p_0(\mc)$ is trivial as we are simulating our $\mc$ posteriors directly with a uniform prior on $[0.4,4.4] \msun$. In principle, $p(\fpeak)$ can be obtained from BayesWave by sampling from the prior (i.e., setting the likelihood to a constant) and then following the same procedure as above to extract $\fpeak$ from the signal reconstruction samples. However, there is a subtlety here: due to correlations between $\fpeak$ and other parameters $\alpha$ in BayesWave, the marginal prior will not generically be the same as the conditional prior; that is, $p(\fpeak)\neq p(\fpeak|\alpha)$. In such cases, the formally correct approach would be to extend the framework of Eq.~\eqref{eqn:finalpost_propto} to account for the other parameters (as for example should be done for the joint inference between EoS and mass distribution discussed above). We defer such an approach to future work and here instead continue with Eq.~\eqref{eqn:finalpost_propto} where $p(\fpeak)$ is the conditional prior, which we obtain through successive BayesWave analyses of pure detector noise. The resultant prior distribution is shown in Fig. \ref{fig:fpeak-prior}.

\subsection{Simulated Post-merger Waveforms}\label{waveforms}

\begin{figure}
    \centering
    \includegraphics[width=0.95\linewidth,trim=0cm 0cm -1cm 0cm]{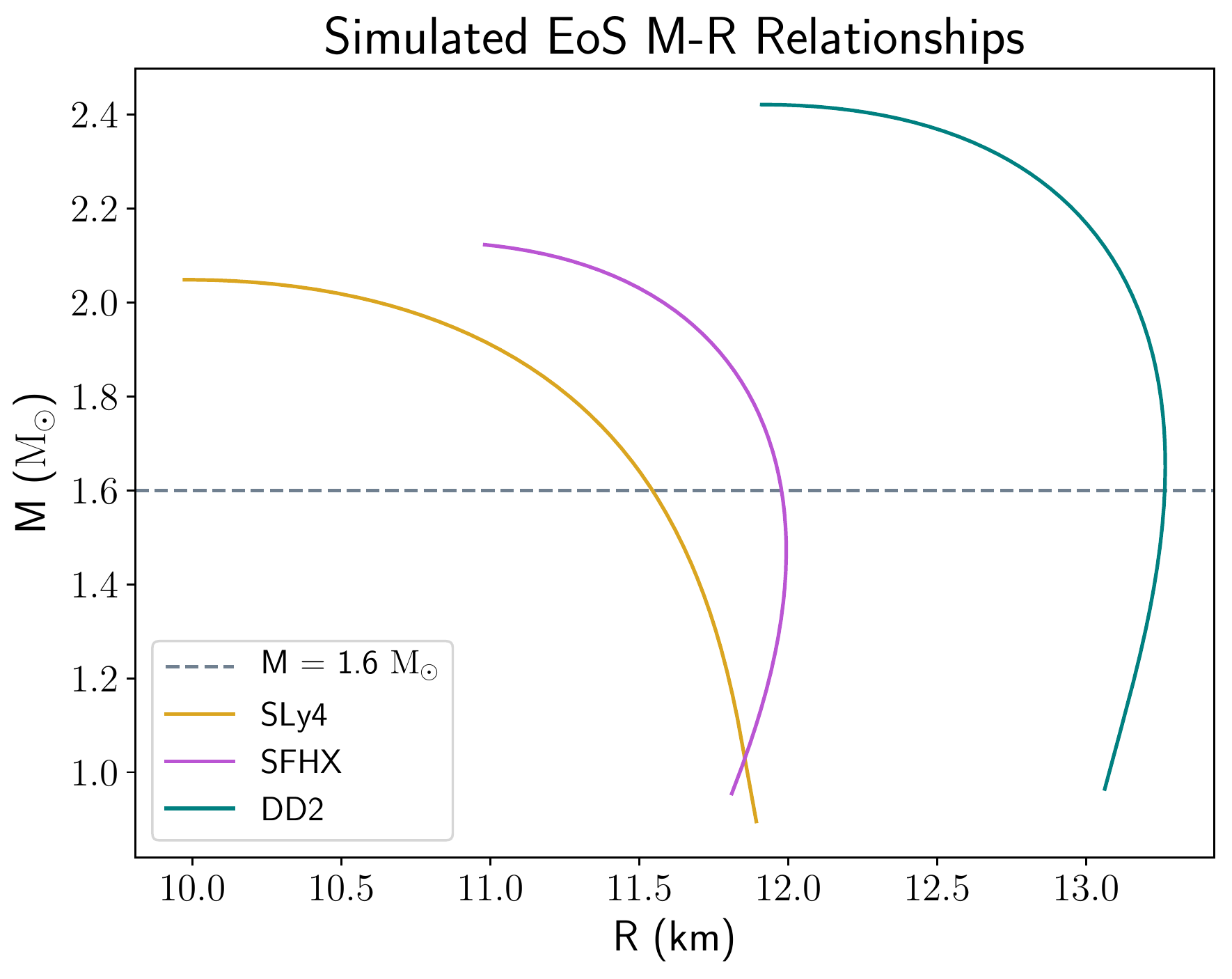}
    \caption{Neutron star mass-radius relationship for each of the equations of state used in this work, along with each relationship's intersection with $\text{M} = 1.6 \msun$. The corresponding values of $\R$ are 11.54 km (SLy4), 11.98 km (SFHX), and 13.26 km (DD2).}
    \label{fig:EoS_MR_rel}
\end{figure}

\begin{figure*}
    % \centering
    \includegraphics[width=0.95\linewidth,trim=0cm 0cm -0.2cm 0cm]{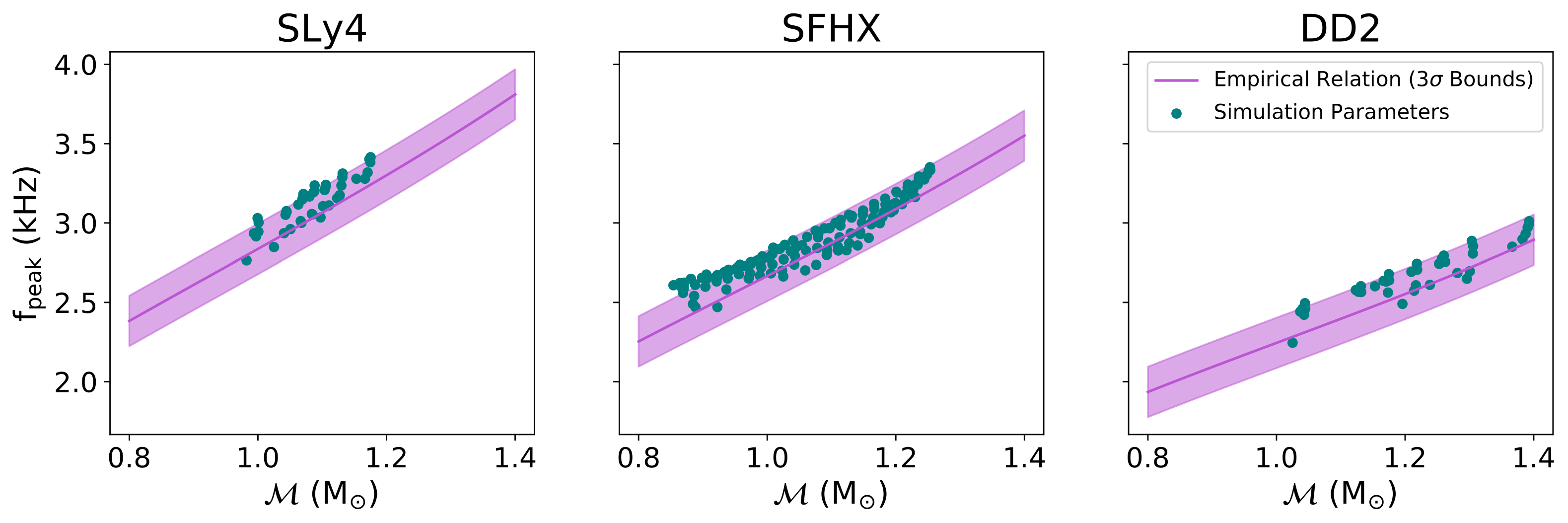}
    \caption{Simulation parameters (post-merger peak frequency $\fpeak$ and chirp mass $\mc$) for each EoS, with the corresponding \citet{vretinaris_empirical_2020} empirical relation with $3\sigma$ bounds for that EoS's value of $\R$.}
    \label{fig:sim_params_all}
\end{figure*}

In this study we use a library of simulated post-merger signals created by relativistic hydrodynamical simulations based on smooth particle hydrodynamics (SPH). For code details we refer to \citep{oechslin_conformally_2002,oechslin_relativistic_2007,bauswein_discriminating_2010}, and note that in this work we employ a Wendland kernel function (see \citep{schaback_kernel_2006,rosswog_sph_2015}) as compared to a cubic spline kernel \citep{monaghan_smoothed_1992} used in previous studies with this code. The new SPH kernel features less noise and numerical damping but does not strongly affect the GW frequency. It leads to slightly tighter peaks in the post-merger spectrum.

This library is comprised of simulations for three different EoSs: DD2 \citep{hempel_statistical_2010}, SLy4 \citep{douchin_unified_2001}, and SFHX \citep{steiner_core-collapse_2013}, representative of stiffer, softer, and medium EoSs, respectively. Softer EoSs result in smaller NS radii and in higher post-merger GW frequencies at the same total binary mass. The DD2, SLy4, and SFHX EoSs were chosen accordingly to explore the effectiveness of our analysis throughout the EoS parameter space. The DD2 \citep{hempel_statistical_2010} and SFHX \citep{steiner_core-collapse_2013} EoSs are derived using the relativistic mean field approximation, whereas SLy4 \citep{douchin_unified_2001} is based on a Skyrme interaction. Mass-radius relationships for each of these EoSs can be found in Fig.~\ref{fig:EoS_MR_rel}. While the DD2 and SFHX models provide the temperature dependence of the EoS, the SLy4 is only available at zero temperature assuming neutrinoless beta-equilibrium. Thus, the latter EoS model is supplemented by an approximate thermal treatment choosing $\Gamma_\mathrm{th}=1.75$ (see \citet{bauswein_testing_2010} for details and a justification of this choice).

For each EoS, we perform simulations for a grid of total binary masses $M_\mathrm{tot}$ and mass ratios $q$. We vary the total binary mass $M_\mathrm{tot}$ in a broad range, but we do not consider systems which promptly form a black hole after merging. Hence, we only study binaries with a total mass well below the threshold mass $M_\mathrm{thres}$ for direct black hole formation. In all of these systems the remnants survive for at least 15~ms before collapsing to a black hole. 
Generally, $M_\mathrm{thres}$ depends on the EoS and exhibits an additional (mild) dependence on the mass ratio $q$ (see e.g. \cite{bauswein_systematics_2021}). Table~\ref{tbl:Mthres} provides the values of $M_\mathrm{thres}$ for the respective EoSs included in this work for mass ratios $q\in\{0.7,0.85,1.0\}$ \cite{bauswein_systematics_2021}. We consider six binary mass ratios $q\in\{0.7, 0.8, 0.85, 0.9, 0.95, 1.0\}$. For the SFHX models, we use a finer $M_\mathrm{tot}$ grid, allowing for a deeper exploration of the analysis using a large number of events for one EoS. In total, the library contains 41 DD2 waveforms, 132 SFHX waveforms, and 41 SLy4 waveforms. The $\mc$ and $\fpeak$ values for all simulations can be found on Zenodo at \url{10.5281/zenodo.7007630}, and are shown  in Fig.~\ref{fig:sim_params_all} along with the empirical relation $f_\mathrm{peak}({\mc})$ of \citet{vretinaris_empirical_2020} for a constant $R_{1.6}$ fixed for the respective EOS.

\begin{table}
    \centering
    \begin{tabular}{lccc}
         EoS & & $M_{\textrm{thres}}\ (\msun)$ & \\
          & $q=0.7$ & $q=0.85$ & $q=1.0$\\
         \hline
         SFHX & 2.925&		    2.975	&	     2.975\\
         DD2 & 	3.275&		    3.325		 &    3.325\\
         SLy4 &	2.775&		    2.825		  &   2.825\\
    \end{tabular}
    \caption{Threshold masses for direct collapse to a black hole at multiple values of the mass ratio $q$ for each EoS in this work.}
    \label{tbl:Mthres}
\end{table}

\subsection{Implementation: the BAYESTACK  Package}\label{code}

The final $\R$ posterior is one-dimensional due to our marginalization over $\fpeak$ and $\mc$. As we utilize KDEs to model the $\fpeak$ and $\mc$ posterior samples for each event, the low dimensionality of this problem allows us to directly compute the final $N$-event posterior on a grid, as opposed to more time-consuming stochastic sampling techniques. Accordingly, we present the hierarchical BAYEsian STACKing (\textsc{bayestack)} Python package as a direct numerical implementation of the formalism described in \S\ref{formalism}. This code is publicly available and can be found on GitHub at \href{https://github.com/criswellalexander/bayestack}{https://github.com/criswellalexander/bayestack}. The central algorithm of the \textsc{bayestack} package as employed in this work is described in brief below.

\begin{enumerate}
    \item A dictionary is created to store the data for $N$ events. Each event is then assigned its own dictionary to hold its associated data and distributions.
    \item For each event, the peak frequency posterior samples are loaded. We then use the PESummary \citep{hoy_pesummarypesummary_2021} reflection-bounded KDE adaptation of the SciPy \citep{virtaneN_scipy_2020} Gaussian KDE \texttt{scipy.stats.gaussian\_kde} with a KDE bandwidth of 0.15 to estimate the $\fpeak$ posterior probability density on [1.5,4] kHz.
    \item A similar procedure is followed for the chirp mass, simulating samples from a normal distribution as described in \S\ref{mchirp-approx} \emdashb or loading pre-computed $\mc$ samples if desired \emdashb and using a KDE to estimate the posterior probability density. 
    \item The pre-computed $\fpeak$ prior samples are loaded (see \S\ref{bayeswave}, \S\ref{priorchoice}, and Fig. \ref{fig:fpeak-prior}). The prior is fit with a \textsc{PESummary} reflection-bounded KDE.
    \item The individual event likelihoods are then computed. For every point in $\R$, the integrands of Eq.~\eqref{eqn:finalpost_propto} are calculated on a grid in $\mc$ with $\Delta\mc=0.005\msun$. For each grid point:
    \begin{itemize}
        \item The original prior $p_0(\mc)$, prior $p(\mc)$, and posterior $p(\mc|D_{\textrm{IN}})$ probabilities for $\mc$ are computed and stored.
        \item The corresponding value of $\hfpij$ is computed for each of $N_{\mathrm{R}}=1000$ sampled $F_{E,j}$ empirical relations.
        \item If $\hfpij$ is in-band (see \S\ref{formalism}), the estimated $\fpeak$ prior and posterior probabilities at that value are stored. 
        \item If $\hfpij$ is out-of-band, the $\fpeak$ likelihood is taken to be 1 as discussed in \S\ref{formalism}.
    \end{itemize}
    \item The grid is then summed over $F_{E,j}$ and $\mc$ to yield the individual marginal likelihood in $\R$, $p(D_i|\R)$.
    \item We then calculate the $N$-event likelihood $\prod_{i=1}^N p(D_i|\R)$ given in Eq.~\eqref{eqn:finalpost_propto}. In addition to calculating the product across all event likelihoods, each $p(D_i|\R)$ is also stored individually. 
    \item This final likelihood is then multiplied by an $\R$ prior of our choice to yield the final posterior $p(\R|D_{i...N})$. We use either a uniform prior on [9,15] km or the multimessenger prior of \citet{huth_constraining_2022} as described in \S\ref{priorchoice}. 
\end{enumerate}

\section{Results}\label{results}

\subsection{Model Validation}\label{sanity}

We perform a set of tests under ideal conditions to ensure the analysis works as intended. For each of these tests, we follow the above procedure, save that we directly simulate the $\fpeak$ posteriors, so as to have fine control over the embedded signal strength. Additionally, this choice allows us the flexibility to consider a broad range of merger masses and EoSs beyond the bounds of our library of simulated post-merger waveforms. For each event, $\mcit$ is determined by drawing $m_1$ and $m_2$ from the distributions used in \citet{abbott_prospects_2020-1} and \citet{petrov_data-driven_2022} (as we use the events of the latter work in our simulations of future observing runs; see \S\ref{obs_run_sims}). 

To simulate $\fpeak$ posterior samples, we draw a mixture of $N_s$ signal and $N_n$ noise samples. The noise samples are drawn from our $\fpeak$ prior samples. To generate signal samples, we first use the associated $\mci$ of each event to determine the corresponding value of $\fpit$, the intrinsic true peak frequency of the event. We do so by using the relation of \citet{vretinaris_empirical_2020}, with dispersion reflective of the relation's intrinsic scatter, $\sigma_{\mathrm{ER}} = 0.053$ kHz. That is, we draw each event's $\fpit$ from $\mathcal{N}(\mu=\hfp,\sigma=\sigma_{\mathrm{ER}})$, with $\hfp = F_E(R_{\mathrm{true}},\mci)$.  Unless otherwise stated, we have (arbitrarily) chosen $R_{1.6,\mathrm{true}}=12.5 $ km for all tests in this section. With the intrinsic $\fpit$ in hand, we then draw signal samples for each event from $\mathcal{N}(\mu=\mu_{f,i},\sigma=\sigma_f)$, where $\sigma_f = 0.1$ kHz and $\mu_{f,i}$ is itself drawn $\mathcal{N}(\mu=\fpit,\sigma=\sigma_f)$ to represent scatter due to detector noise. The posterior width $\sigma_f$ is reflective of low-SNR BayesWave $\fpeak$ recoveries (see, e.g., the lower panel of Fig.~\ref{fig:BW-weak}.)

The result of this process is a set of simulated $\fpeak$ posteriors for which the ratio of signal samples to noise samples $N_s:N_n$ acts as a loose proxy for signal SNR. A posterior with $N_s=0$ contains no hint of a signal whatsoever; conversely, one with $N_n=0$ would be emblematic of a tight, normally-distributed posterior about the injected $\mu_f$. One can then move along the spectrum created by these end-points to simulate a variety of ambiguous $\fpeak$ posteriors of differing strengths, representative of the $\fpeak$ posteriors produced by BayesWave for different post-merger signal SNRs. However, we stress that $N_s:N_n$ is only a proxy for SNR insofar as increasing $N_s:N_n$ correlates with increasing SNR; the two are not interchangeable, nor can one directly convert between them.

\begin{figure}
    \centering
    \includegraphics[width=1\linewidth]{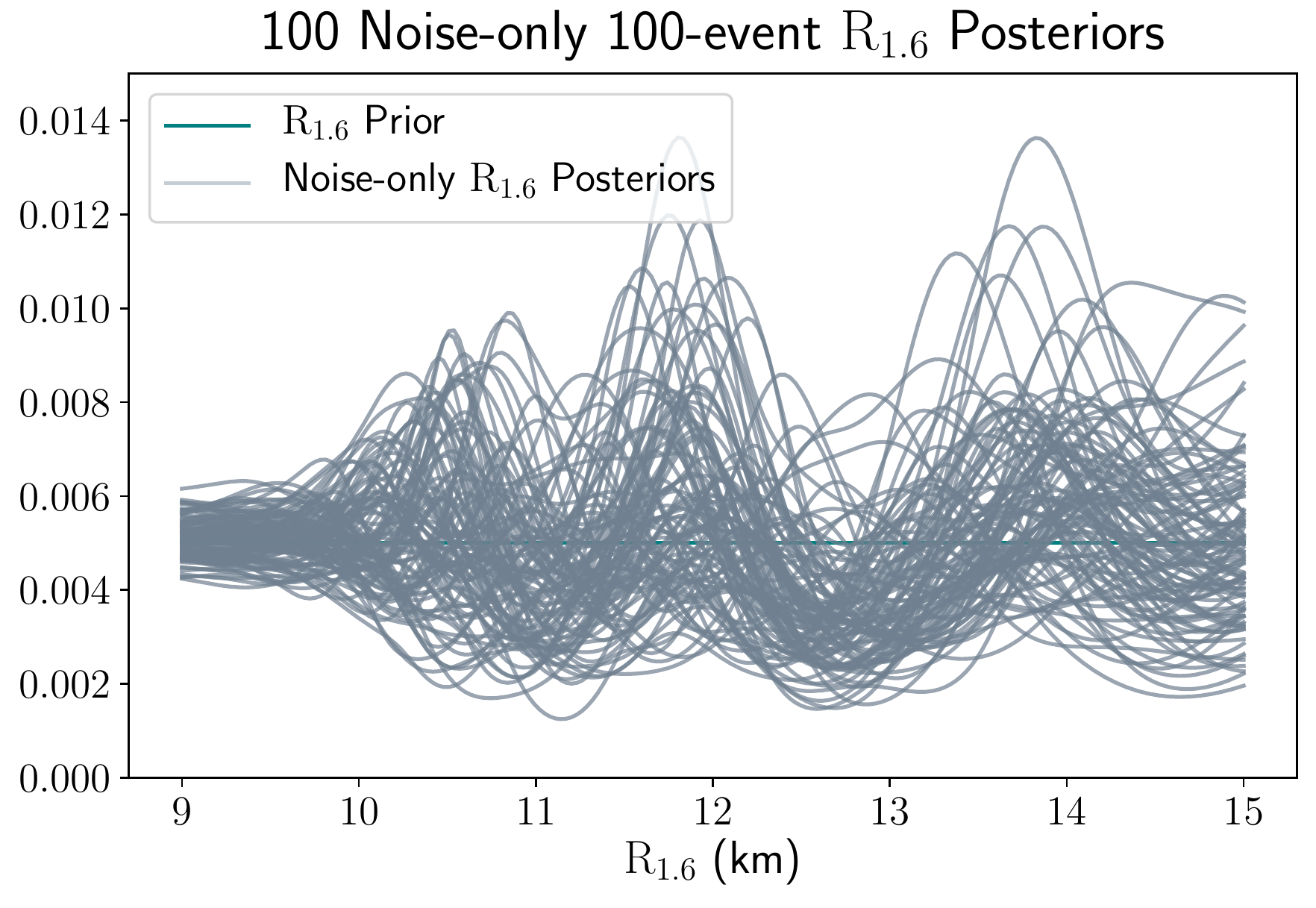}
    \caption{100 realizations of $\R$ posteriors arising from 100 events consisting solely of noise draws. These distributions are representative of what can be produced from random fluctuations in the $\fpeak$ posteriors, absent any impact from an embedded signal. As expected, while we do observe some level of fluctuation in these $\R$ posteriors, the set as a whole is consistent with the prior. The flattening at low $\R$ arises from our band-limited search; as discussed in \S\ref{formalism}, most predicted values of $\hfp$ for $\R\lesssim10$ km fall out-of-band and do not contribute information to our final posterior. As a result, most analyses without a significant signal present in the data will tend to cleave to the prior for these values of $\R$.}
    \label{fig:MV1_noiseonly}
\end{figure}

We begin by considering the case of uninformative $\fpeak$ data, with $N_s=0$. We apply the analysis to a set of 100 synthetic ensembles, each containing 100 events, whose $\fpeak$ posteriors consist solely of noise samples ($N_s=0, N_n=5000$). As no EoS information is encoded in the $\fpeak$ data, we expect to recover results consistent with our prior on $\R$, with some fluctuations due to random alignment across events of noise samples with the empirical relations. As is shown in Fig. \ref{fig:MV1_noiseonly}, this is what occurred. Moreover, this set of $\R$ posteriors from uninformative data provides us with a reference dataset with which to compare our recoveries of informative data.

\begin{figure}
    \centering
    \includegraphics[width=1\linewidth]{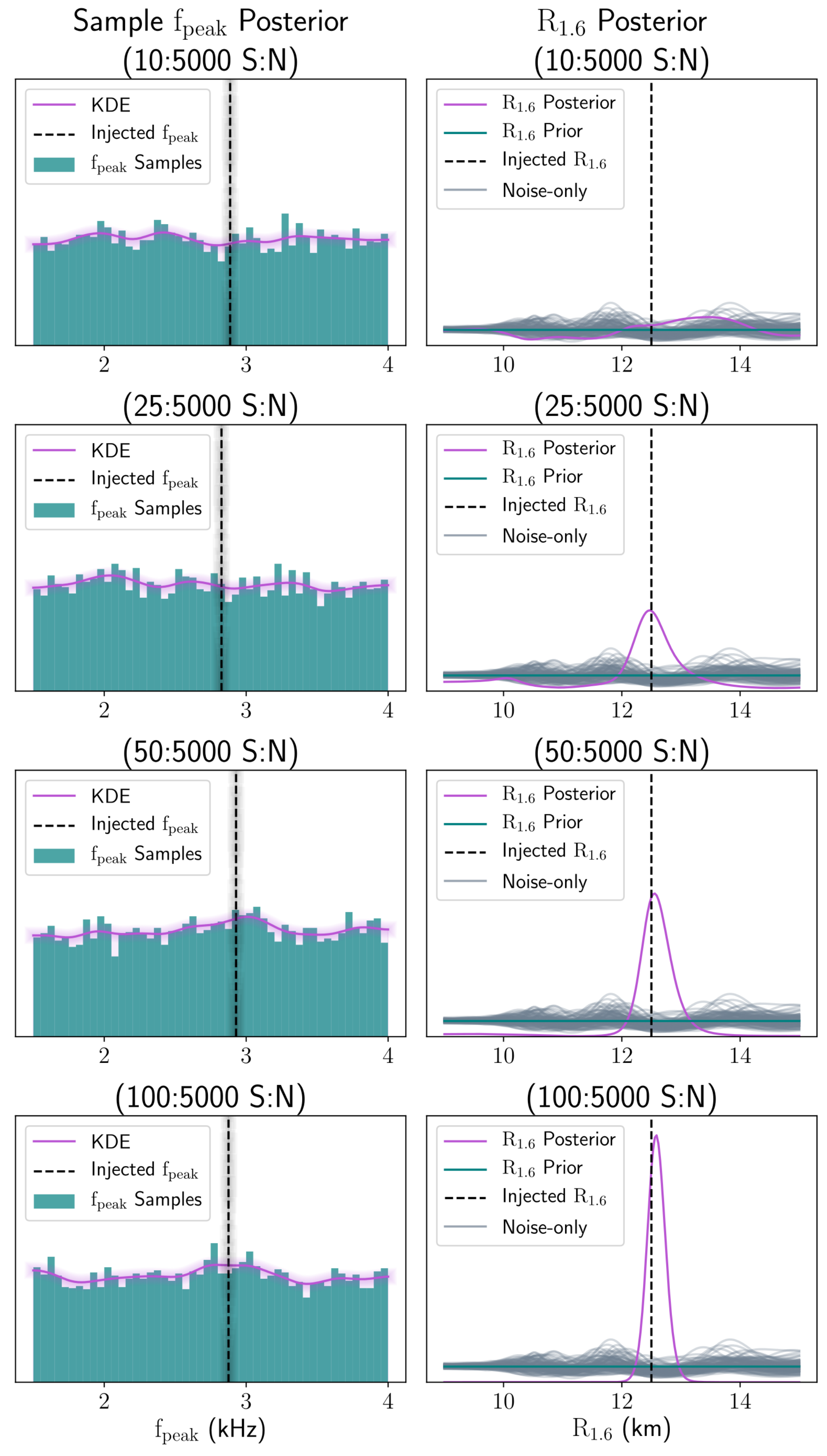}
    \caption{Sample synthetic $\fpeak$ posteriors and corresponding 100-event ensemble $\R$ posteriors for increasing ratio of signal samples to noise samples (S:N). Note that in all cases, $N_s$ is sufficiently small that the corresponding sample $\fpeak$ posterior is not strongly peaked (and in fact by eye appears to be entirely uninformative). For reference, we have also plotted the set of noise-only $\R$ posteriors described in \S\ref{sanity}. From top to bottom, these posteriors correspond to $\R$ means and 95\% C.I. of $12.3^{+2.5}_{-3.2}$ km, $12.1^{+2.1}_{-2.9}$ km, $12.5^{+0.9}_{-2.6}$ km, and $12.59^{+0.33}_{-0.31}$ km, respectively.}
    \label{fig:MV2_10panel}
\end{figure}

We then turn to the most pertinent test case: an ensemble of weak signals. With our same 100-event ensemble, we slowly increase the ratio of $N_s:N_n$ and investigate how the analysis responds to increasing signal strength in the weak-signal regime. In Fig. \ref{fig:MV2_10panel} we show sample simulated $\fpeak$ posteriors and the corresponding final $\R$ posterior for the 100-event ensemble for each ratio of $N_s:N_n$.

At low $N_s:N_n$, we obtain an $\R$ posterior consistent with random fluctuations due to noise. As $N_s:N_n$ increases, we rapidly accrue posterior density around the injected value of $\R=12.5$ km. By $N_s=25$, the posterior peak begins to emerge from the noise-only distributions, by $N_s=50$, the peak is unmistakable, and by $N_s=100$, we are able to exclude wide swathes of the parameter space, corresponding to a mean and 95\% C.I. of $\R=12.59^{+0.33}_{-0.31}$ km.

\begin{figure}
    \centering
    \includegraphics[width=1\linewidth]{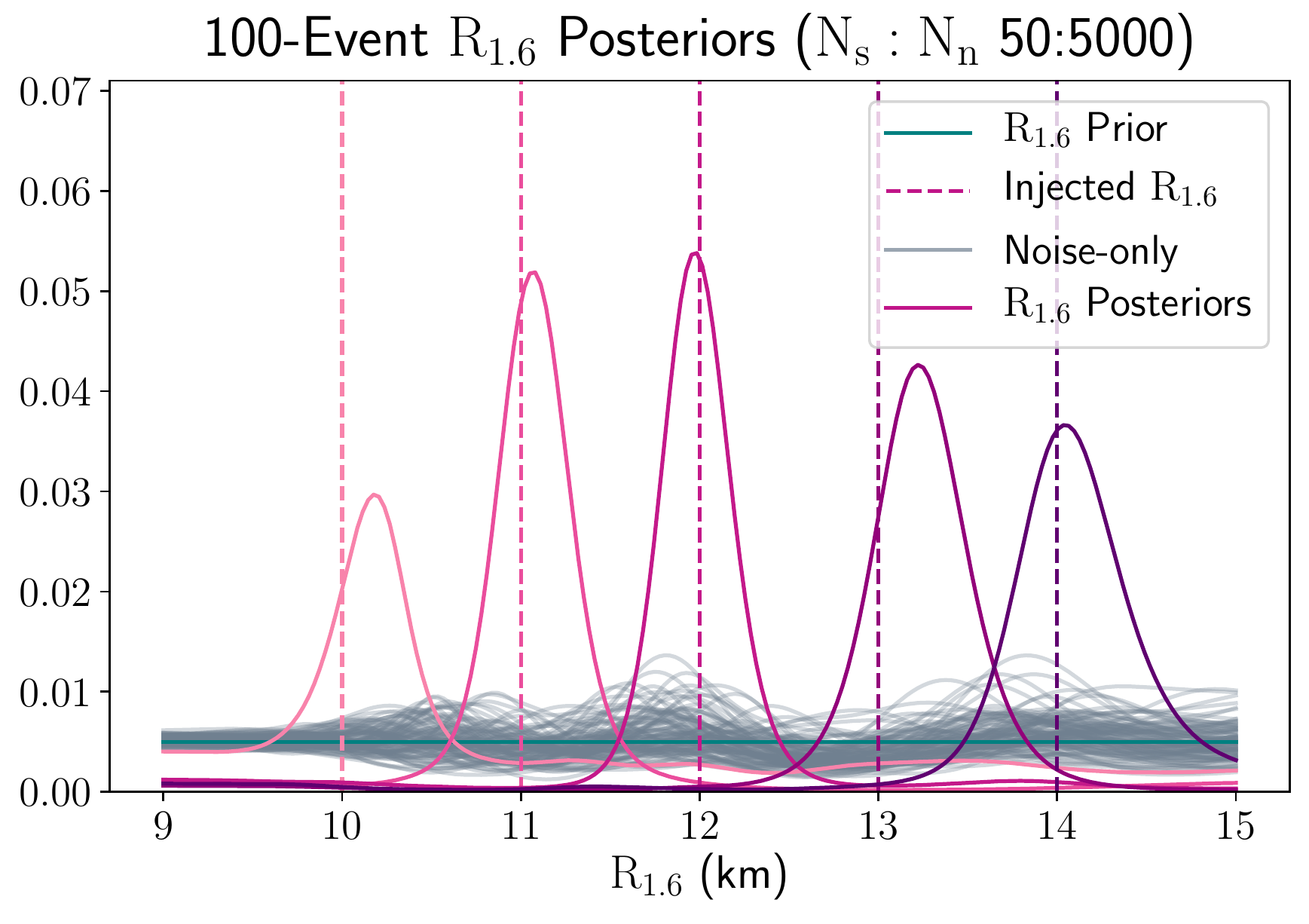}
    \caption{100-event ensemble $\R$ posteriors with $N_s=50, N_n=5000$ for different injected values of $\R$. For reference, we have also plotted the set of noise-only $\R$ posteriors shown in Fig. \ref{fig:MV1_noiseonly}. We accrue posterior density around the injected value in each case, although the recovery strength is reduced by mitigating factors at low and high $\R$.}
    \label{fig:MV3_multiR}
\end{figure}

We then explore the efficacy of our analysis across the $\R$ parameter space. We repeat the 100-event, ($N_s=50,N_n=5000$) test, changing the injected value of $\R$ for each iteration. As shown in Fig. \ref{fig:MV3_multiR}, we are able to recover the injected value in each case, but the strength and precision of the recovery has some dependence on $\R$. This behavior is largely driven by two factors. At low $\R$, $\fpeak$ tends to be higher, and more frequently falls outside of our analysis frequency band. As fewer signals are contained within our data, less overall information is available and the resulting posterior is correspondingly weaker. At high $\R$, however, the increased posterior width is simply driven by a larger intrinsic scatter in the empirical relation. As can be seen in Fig. \ref{fig:ER_dist}, as we approach higher values of $\R$, the empirical relation curves in the $\fpeak-\mc$ plane become shallower and closer together for equal intervals in $\R$. Accordingly, the ability of a ($\fpeak,\mc$) pair to differentiate between nearby values of $\R$ is reduced.

\begin{figure}
    \centering
    \includegraphics[width=1\linewidth]{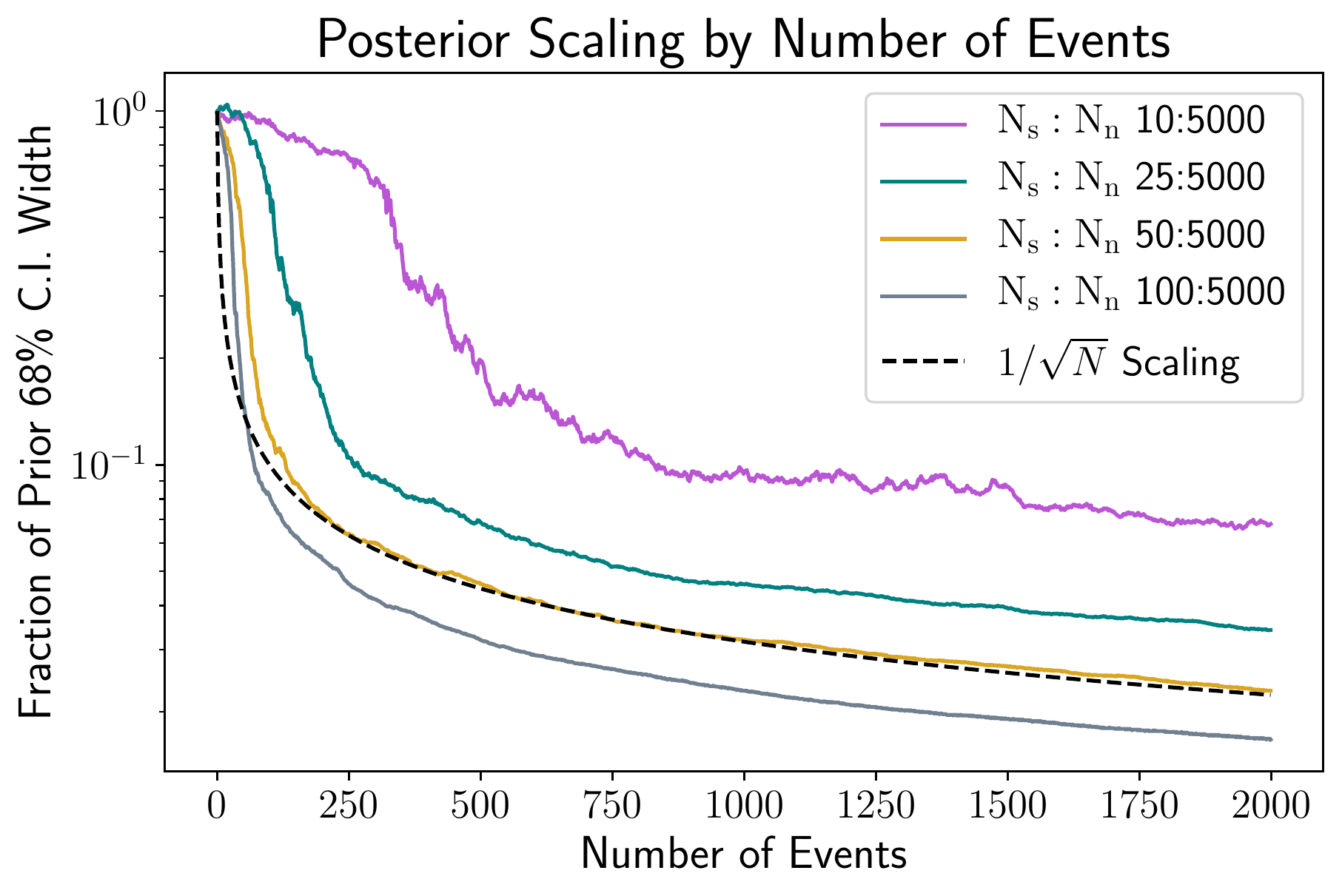}
    \caption{Scaling with $N_{\mathrm{events}}$ of the ensemble $\R$ posterior 68\% C.I. width as a fraction of the prior C.I. width for a 2000-event ensemble with $\R=12.5$ km and four different injected signal strengths. Note that the same behavior occurs for other C.I. percentages, but viewing the entire trend at low signal strengths for e.g. 95\% C.I. requires many more events.}
    \label{fig:post_scale_2k}
\end{figure}

Finally, we investigate the efficacy scaling of our analysis with the number of events included. As in Fig.~\ref{fig:MV2_10panel}, we consider $N_s:N_n$ ratios of $10:5000$, $25:5000$, $50:5000$, and $100:5000$ with $\R=12.5$ km and analyze a 2000-event ensemble generated as described above. Fig. \ref{fig:post_scale_2k} shows the evolution with increasing $N_{\mathrm{events}}$ of the ensemble posterior 68\% C.I. width evolution as a fraction of the original prior C.I.. In each case, we can see three stages of posterior evolution (although different stages dominate for different signal strengths). For low $N_{\mathrm{events}}$, few gains are made. However, as the ensemble becomes more robust and covers more of the parameter space, the power of our analysis comes to the forefront, and we enter a second phase where the posterior drastically and rapidly improves. Ultimately, the diversity of events saturates, and most additional events do not explore a new area of the parameter space; accordingly, the evolution moves into its final stage, wherein its slope matches that of the ``standard" $1/\sqrt{N}$ evolution.

We emphasize that, for each case considered, the individual $\fpeak$ posteriors remain ambiguous, but our analysis is able to combine the trace amounts of information embedded therein to produce informative constraints on $\R$. This showcases the power of our hierarchical approach under ideal, controlled conditions. We now turn to investigate how our analysis performs on its intended input: the peak frequency posteriors of BayesWave.

\subsection{Simulations of Future Observing Runs}\label{obs_run_sims}
We perform simulations of the next two LVK observing runs (O4 and O5), using the catalogues of \citet{petrov_data-driven_2022}. Originally created to accurately explore the prospects for EM followup of GW events in O4/O5, these catalogues comprise realistic simulations of compact binary merger detections in O4 and O5, using astrophysical distributions of binary parameters and SNR detection thresholds reflecting those utilized in O3. For BNS mergers specifically, NS component masses $m_1$ and $m_2$ were normally distributed as $\mathcal{N}(1.33\msun,0.09\msun)$ and component spins were uniformly distributed on [0,0.05], following \citet{abbott_prospects_2020-1}. Events were isotropically distributed in space, given redshifts uniformly distributed in comoving rate density, and generated with the rate measured in the latest O3 results for the BNS astrophysical rate density of $320^{+490}_{-240}\; \textrm{Gpc}^{-3}\textrm{yr}^{-1}$ (see \citet{petrov_data-driven_2022} for further details).

\begin{figure*}
    \centering
    \includegraphics[width=1\linewidth]{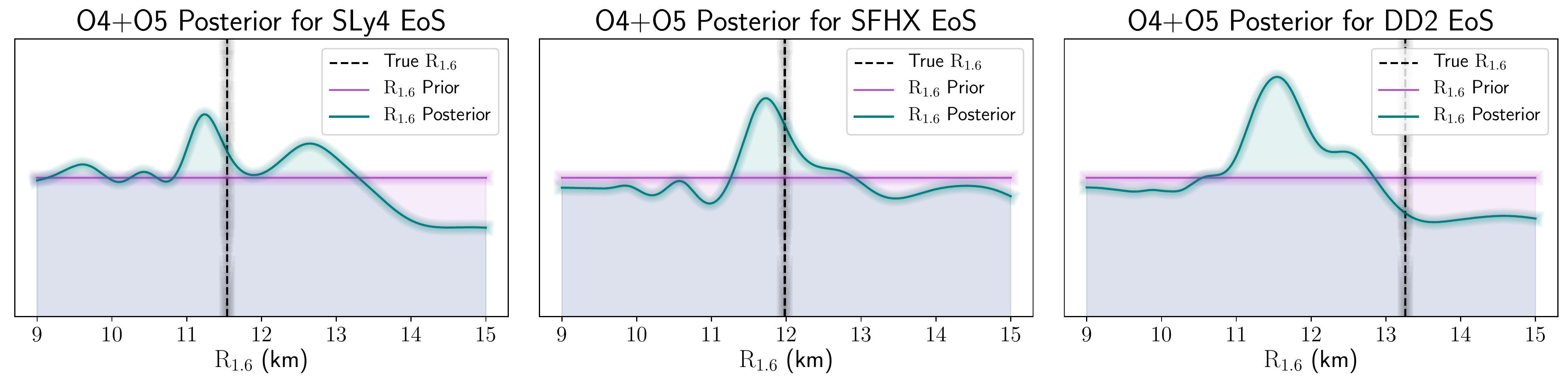}
    \caption{$\R$ posteriors for simulations of LVK's upcoming O4 and O5 observing runs for the SLy4, SFHX, and DD2 equations of state. These posteriors correspond to means and 95\% C.I. of $\R=11.8^{+3.0}_{-2.6}$ km, $\R=12.0^{+2.8}_{-2.8}$ km, and $\R=11.8^{+3.0}_{-2.7}$ km, respectively, as compared to the original uniform prior mean and 95\% C.I. of $\R=12.0^{+2.9}_{-2.9}$ km. While there appear to be hints of a signal in the SLy4 and SFHX cases, the final posterior is not conclusive and could arise from noise fluctuations, as seen in the case of DD2.}
    \label{fig:O4O5_post_3eos}
\end{figure*}

\begin{figure*}
    \centering
    \includegraphics[width=1\linewidth]{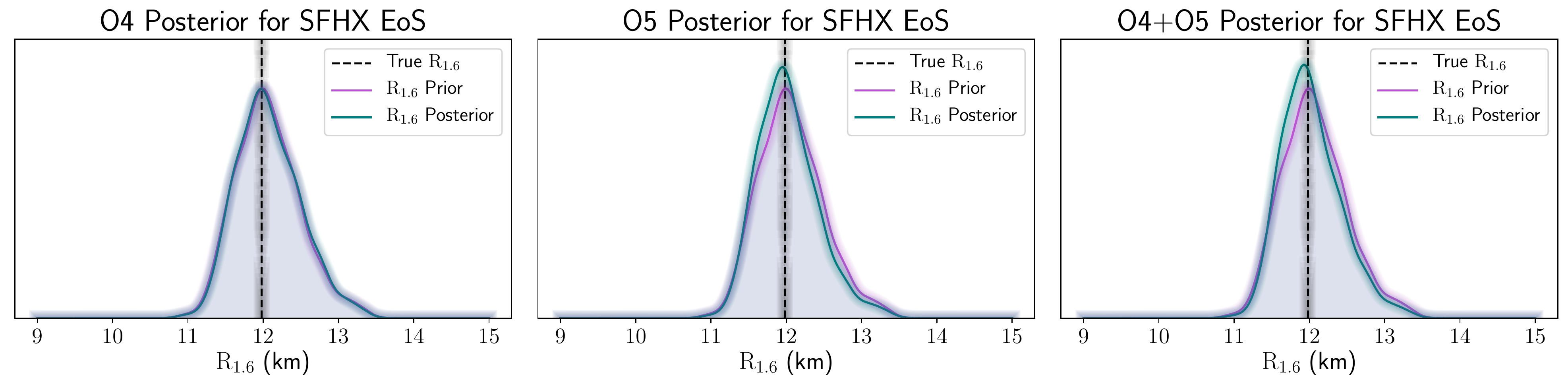}
    \caption{$\R$ posteriors for simulations of LVK's upcoming O4 and O5 observing runs with the SFHX EoS and the multimessenger prior of \citet{huth_constraining_2022}. The individual O4 and O5 posteriors are presented along with the combined O4+O5 posterior. For O4, O5, and O4+O5, these posteriors correspond to
    means and 95\% C.I. of $\R=12.07^{+0.95}_{-0.75}$ km, $\R=12.02^{+0.95}_{-0.70}$ km, and $\R=12.02^{+0.93}_{-0.68}$ km, respectively. For comparison, the prior mean and 95\% C.I. are $\R=12.07^{+0.98}_{-0.77}$ km. The combined O4+O5 posterior provides a reduction of the 95\% C.I. width by $0.13$ km (7.5\%) over the prior.}
    \label{fig:SFHX_O4O5_post}
\end{figure*}

We use a one year duration for both O4 and O5. \citet{petrov_data-driven_2022} predict a median annual number of detected BNS mergers of 34 for O4 and 190 for O5 with an SNR cut of inspiral SNR$>$8, reflective of the alerts sent out during O3; see \citet{petrov_data-driven_2022} for justification. We therefore take 34 and 190 draws from the \citet{petrov_data-driven_2022} distributions to simulate O4 and O5, respectively, ensuring that both realizations are representative of the median number of events in the nearby universe (which will be the most important for post-merger searches). The \citet{petrov_data-driven_2022} simulations contain many BNS mergers fated to direct collapse, but we are only interested in the HMNS post-merger scenario. Accordingly, we discard all events with $M_{\mathrm{tot}} > 3 \msun$. Such a hard cut is somewhat inelegant; for example, $M_{\mathrm{thres}}$ for the DD2 EoS lies above this cut by $\sim0.3\msun$ whereas that of SLy4 lies below it by $\sim0.15\msun$, leading us to discard events we should keep in the former case and vice versa for the latter. Ultimately, this approach is a necessary evil; it allows for identical treatment and therefore direct comparison of each EoS with respect towards the others, and as such suits our present purposes. A more refined treatment of $M_{\mathrm{thres}}$ and the direct collapse scenario is left to future work (see \S~\ref{future}). After this cut, we are left with 26 (83) events in O4 (O5). For each selected event in the catalogue, we analyze the post-merger waveform from our library closest\footnote{Specifically, we select the waveform with the smallest 2-norm distance in $m_1-m_2$ space from the original catalogue event's parameters.}  to the catalogue parameters, injected and recovered using the event's distance as given in \citet{petrov_data-driven_2022}. The final set of events with associated $\fpeak$ posteriors can be found on Zenodo at \url{10.5281/zenodo.7007630}. We use the detector strain noise spectral density released in LIGO-T2000012 \citep{oreilly_ligo-t2000012-v1_2020}. For the purposes of this study, we assume each event was optimally-inclined and detected by LIGO Hanford, LIGO Livingston, and VIRGO interferometers. For each post-merger signal, we simulate a corresponding chirp mass posterior as discussed in \S \ref{mchirp-approx}. We perform our simulations thrice, once for each EoS (SLy4, SFHX, and DD2). In Fig. \ref{fig:O4O5_post_3eos}, we present the full O4+O5 simulation posterior for each EoS with a uniform prior in $\R$. For SLy4, SFHX, and DD2, the full O4+O5 posteriors correspond to means and 95\% C.I. of $\R=11.8^{+3.0}_{-2.6}$ km, $\R=12.0^{+2.8}_{-2.8}$ km, and $\R=11.8^{+3.0}_{-2.7}$ km, respectively, compared to the original uniform prior mean and 95\% C.I. of $\R=12.0^{+2.9}_{-2.9}$\,km. Additionally, we present the O4, O5, and combined O4+O5 posteriors for the SFHX EoS in Fig. \ref{fig:SFHX_O4O5_post} with the multimessenger prior of \citet{huth_constraining_2022}. The full O4+O5 ensemble posterior equates to a mean and 95\% C.I. of $\R=12.02^{+0.93}_{-0.68}$ km, a reduction of 0.13 km (7.5\%) compared to the original prior 95\% C.I. of $\R=12.07^{+0.98}_{-0.77}$ km. As would be expected, the final posterior is primarily driven by O5 events, reflecting both the increased detector sensitivity and event count. However, the combined O4+O5 posterior does improve slightly when compared with O5 alone, indicating that the O4 data is in fact providing some information.

\begin{figure*}
    \centering
    \includegraphics[width=0.9\linewidth]{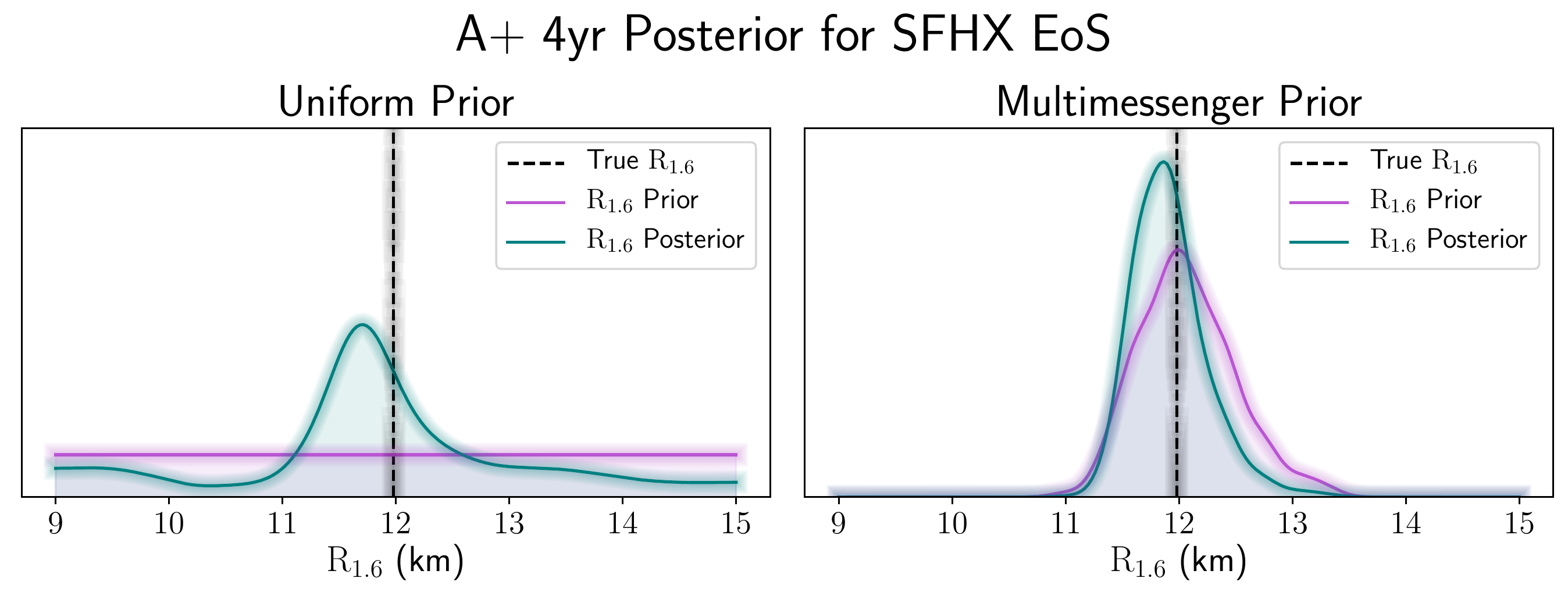}
    \caption{$\R$ posteriors arising from a simulation of 4 years of operation at A+ sensitivity for the SFHX EoS, presented with both a uniform prior and the multimessenger prior of \citet{huth_constraining_2022}. For the multimessenger (uniform) prior, we attain posterior means and 95\% C.I. of $\R=11.91^{+0.80}_{-0.56}$ km ($\R=11.9^{+2.7}_{-2.6}$ km). For reference, the multimessenger (uniform) prior means and 95\% C.I. are $\R=12.07^{+0.98}_{-0.77}$ km ($\R=12.0^{+2.9}_{-2.9}$ km). In the case of the multimessenger prior, we achieve a reduction in the 95\% C.I. width of $0.39$ km (22\%).}
    \label{fig:Aplus4yr}
\end{figure*}

While these posteriors do not comprise stringent constraints on $\R$, there is an accumulation of posterior probability about the simulated values of $\R$ corresponding to the SFHX and SLy4 EoSs. However, the result for the DD2 EoS is less informative and does not show accumulation about the true value as seen with the other two equations of state. That we see such behavior in the case of DD2 is not entirely unsurprising; as demonstrated in Fig. \ref{fig:MV3_multiR}, our analysis is most sensitive for middling values of $\R$, and decreases in statistical power towards the high-$\R$ regime occupied by the DD2 EoS. The nature of the O4+O5 posteriors is consistent with the behavior of our analysis as seen in \S\ref{sanity} for the case of low signal strength and (relatively) small $N_{\mathrm{events}}$; it is possible that the embedded information contained in the O4+O5 simulation data is sufficiently weak that this high-$\R$ reduction in sensitivity proved insurmountable. Repetition of the DD2 O4+O5 study with all events injected at $20$ Mpc \emdashb corresponding to a mix of low- and mid-SNR signals \emdashb yielded posterior accumulation about the injected value of $\R=13.26$ for DD2, lending additional credence to this explanation. We further investigate methods for characterizing the probability of the observed posterior accumulation arising from noise alone in \S\ref{mcvalidation}.

As seen in \S\ref{sanity}, the primary source of our resolving power in the low-SNR regime is the inclusion of additional events. If the posterior accumulation observed in e.g. the SFHX O4+O5 simulation is reflective of real information contained within the post-merger data, we would expect to see continued gains as we increase $N_{\mathrm{events}}$. Accordingly, we consider what may be possible with 4 years of data from the LIGO/Virgo network operating at O5 (A+) sensitivity. We follow the procedure described for the simulations of O4 and O5, save that we draw 760 events (4 years at a median annual detection count of 190 events), of which 357 are retained after our cut in $M_{\mathrm{tot}}$. We consider only the SFHX EoS, as the finer $q-M_{\mathrm{tot}}$ grid used to compute the SFHX waveforms lends itself to deep studies with larger $N_{\mathrm{events}}$. The resulting combined posterior can be found in Fig. \ref{fig:Aplus4yr}, and is presented both with a uniform prior and with the astrophysical prior of \citet{huth_constraining_2022} as a demonstration of what this method may be able to contribute to our knowledge of the NS EoS in concert with current constraints. For the multimessenger prior, the final $\R$ posterior equates to a mean and 95\% C.I. of $\R=11.91^{+0.80}_{-0.56}$ km, a reduction in the 95\% C.I. width of $0.39$ km (22\%) over the prior 95\% C.I. of $\R=12.07^{+0.98}_{-0.77}$ km. We emphasize that this posterior arises solely from post-merger signals whose individual $\fpeak$ posteriors are consistent with nondetection; this result demonstrates the potential of our approach to produce informative constraints on the NS EoS from post-merger GWs with current-generation detector sensitivities. 

\subsection{Validation}\label{mcvalidation}
As the post-merger signals considered in this work are individually consistent with nondetection, it is important that we be able to assess if the final posterior for $\R$ has arisen from running our analysis over detector noise, as opposed to containing useful information about the NS EoS. To this end, we repeat the simulations of O4, O5, and the A+ 4 year dataset 1000 times each, following the procedure of \S\ref{obs_run_sims}, save that the $\fpeak$ posteriors for each event do not contain any information about the post-merger signal. Instead, each $\fpeak$ posterior consists only of an equivalent number of prior draws (which in turn are draws from noise-only BayesWave posteriors; see \S\ref{priorchoice}). The result of this process is a set of ``null'' posterior densities for $\R$ for each of our observing run simulations.

This null set of posterior densities is then used to investigate whether we can differentiate our simulation results from the noise-only case. Standard model comparison techniques, such as Bayes factors, are not well-suited to our setting; we therefore present the following method, based on the usual statistical hypothesis testing framework, to characterize the difference between the set of null densities and the final $\R$ posterior distributions we obtain from the original simulations. For each null $\R$ posterior density, we compute two metrics: the Wasserstein distance\footnote{Also called the earth-mover's distance.} from the prior to the posterior and the standard 95\% C.I. posterior width. The Wasserstein distance between two probability density functions $f$ and $g$ is defined as
\[
	W(f,g)=\int_\mathbb{R}|F(x)-G(x)|\,dx,
\]
where $F$ and $G$ are the cumulative distribution functions associated with $f$ and $g$, respectively. The distribution of values for a given metric associated with the collection of null densities forms a simulated null distribution; we can then use this distribution to estimate the p-value of the same metric as computed for the final $\R$ posterior for each of our simulated observing runs. For the Wasserstein distance, we estimate the p-value as the proportion of the null distribution greater than the observed density distance, whereas the equivalent estimate for the 95\% C.I. width is the proportion of null C.I. widths \textit{less} than the C.I. width of the observed density. The metric values and their associated p-values for each EoS and simulation are shown in Tables~\ref{tbl:pvalues_uniform} and~\ref{tbl:pvalues_multimessenger} for analyses with the uniform and multimessenger priors, respectively.
\begin{table}
    \centering
    \begin{tabular}{llcccc}
         Run & EoS & WD & p (WD) & CI95 & p (CI95) \\
         \hline
         O4+O5 & SLy4 & 0.21 & 0.48 & 5.64 km & 0.17 \\
         O4+O5 & SFHX & 0.10 & 0.88 & 5.70 km & 0.42 \\
         O4+O5 & DD2 & 0.24 & 0.38 & 5.64 km & 0.17 \\
         A+ 4yr & SFHX & 0.62 & 0.23 & 5.37 km & 0.14 \\
    \end{tabular}
    \caption{Estimated p-values for the Wasserstein distance (WD) and 95\% credible interval width (CI95) metrics for simulation results with a uniform $\R$ prior.}
    \label{tbl:pvalues_uniform}
\end{table}
\begin{table}
    \centering
    \begin{tabular}{llcccc}
         Run & EoS & WD & p (WD) & CI95 & p (CI95) \\
         \hline
         O4+O5 & SLy4 & 0.02 & 0.97 & 1.78 km & 0.67 \\
         O4+O5 & SFHX & 0.05 & 0.68 & 1.60 km & 0.20 \\
         O4+O5 & DD2 & 0.07 & 0.49 & 1.63 km & 0.24 \\
         A+ 4yr & SFHX & 0.16 & 0.40 & 1.36 km & 0.12 \\
    \end{tabular}
    \caption{Estimated p-values for the Wasserstein distance (WD) and 95\% credible interval width (CI95) metrics for simulation results with the \citet{huth_constraining_2022} multimessenger $\R$ prior.}
    \label{tbl:pvalues_multimessenger}
\end{table}
Despite some posteriors hinting at the presence of a signal by eye, the corresponding p-values for our simulated results are insufficient to claim a decisive constraint. This is reflected in the fact that none of our posteriors confidently exclude large swathes of the $\R$ parameter space, instead displaying in most cases posterior accumulation near the injected value of $\R$. We note that the A+ 4yr dataset has the highest significance for all configurations. The case of the A+ 4yr dataset with the multimessenger prior has the highest overall significance at $p=0.12$ for the 95\% C.I. metric. Both the corresponding analysis with a uniform prior and that of the multimessenger prior and Wasserstein distance metric yield lower significance estimates. The increased significance of the multimessenger result is not entirely surprising; as discussed in \S\ref{sanity}, our analysis is less effective at constraining the far tails of the $\R$ distribution, and gains power accordingly when combined with other data that rules out those tails. The difference in significance for the A+ 4yr result between the 95\% C.I. and Wasserstein distance metrics may indicate a need for further exploration; in general, this kind of approach to model comparison is an interesting statistical problem and warrants further work beyond the scope of this paper.

\section{Discussion and Conclusions}
We present a hierarchical Bayesian approach for constraining the NS EoS through BNS post-merger GW emission. The analysis is able to combine an ensemble of barely-informative, marginal detections of the BNS post-merger signal in a coherent fashion to achieve informative constraints on the NS EoS. Additionally, we provide results of the analysis as applied to simulations of the upcoming LVK observing runs O4 and O5 for three different NS equations of state, alongside an investigation of what could be achieved with an extended run of 4 years at A+ sensitivity. The results of these simulations indicate that this method has the potential to produce informative posteriors on $\R$ from BNS post-merger GW \textit{with current generation detectors.} If reflected in the GW observations of coming years, the resulting constraints would comprise a novel contribution to our understanding of the NS EoS from a previously inaccessible source.

\subsection{Prospects for Future Detection}\label{future-detection}
The results we achieve in simulations at current-generation sensitivity indicate a promising future as detector sensitivity improves. In the near future, the proposed LIGO Voyager upgrades would greatly improve detection and characterization of BNS inspiral signals \citep{adhikari_cryogenic_2020}. In addition to Voyager's own increased high-frequency sensitivity, a promising project by OzGrav, NEMO \citep{ackley_neutron_2020}, is explicitly designed for sensitivity in the high-frequency regime. A 2.5-generation detector network combining BNS inspiral detections by Voyager and post-merger data from NEMO could enable this approach to yield greatly improved constraints on the NS EoS. 

The advent of 3rd-generation detectors such as Einstein Telescope \citep{maggiore_science_2020} and Cosmic Explorer \citep{reitze_cosmic_2019} will bring BNS post-merger GW well within the detection horizon. As we enter the fully-operational era of 3rd-generation detectors, it is expected that we will observe a handful of unambiguous post-merger signals each year. However, it is likely that in the early days of these detectors, a large fraction of any post-merger detections will be marginal. Moreover, we expect to see many more marginal signals of all types than so-called ``gold-plated" detections. This will be no less true of the BNS post-merger signal \citep{haster_inference_2020}, providing utility for our approach well into future generations of GW detectors.

\subsection{Potential to Probe Exotic Equations of State }\label{phase-transitions}

It is worth noting that the large number of BNS detections required to achieve competitive constraints on $\R$ using this method will also yield greatly improved constraints on the NS EoS through GW inspiral measurements of $\Lambda(m)$ and EM observations of kilonovae~\citep{landry_nonparametric_2020,gamba_waveform_2021,pratten_impact_2021,chatziioannou_uncertainty_2022}. However, even if the constraints placed by this method are outpaced by the aforementioned inspiral- and EM-counterpart-derived constraints, there is still significant utility in considering NS EoS constraints specifically arising from post-merger GW. The reasons for this are twofold. First, \textit{any} additional information is useful. Even mild constraints produced through our method will improve the potentially more accurate inspiral measurements. More importantly, though, there is a possibility that some post-merger remnants may undergo a phase transition occurring during the merger. If this is the case, our post-merger analysis will yield EoS constraints that differ from inspiral constraints arising from the same BNS population \citep{bauswein_identifying_2019,wijngaarden_probing_2022}. If the former and latter distributions were to exclude each other to a statistically significant extent, our approach could provide strong evidence for exotic matter in BNS post-merger remnants. This possibility renders crucial the concurrent, comparative consideration of both inspiral and post-merger constraints.

\subsection{Future Work}\label{future}
Looking forward, this work has several natural extensions. First, \citet{vretinaris_empirical_2020} provide empirical relations for several values of $\text{R}_{\text{1.x}}$ beyond $\R$. Extending this analysis to provide constraints throughout the mass-radius relation space would greatly improve its ability to differentiate between proposed equations of state. 

Additionally, given this work's results for current-generation detectors, it is certainly worth investigating what can be accomplished with our analysis when applied to 2.5- and 3rd-generation detectors. Realistic simulations of a 2.5-generation network consisting of Voyager and NEMO may provide a new science case for the implementation of such a network, and a similar treatment of Cosmic Explorer/Einstein Telescope would provide important insight into what can be accomplished in the future \emdashb and how constraints derived from marginal detections can compare to those from unambiguous ones. With the extended reach of these detectors, it will be necessary to directly incorporate inspiral redshift posteriors into our statistical hierarchy. Additionally, a more refined treatment of remnant classification would allow us to reduce the potential impact of contamination from direct collapse scenarios as well as avoid discarding true HMNS remnants with masses near the (unknown) mass threshold for direct collapse, $M_{\text{thres}}$. This could be accomplished through determination of the probability of direct collapse \emdashb marginalized over inspiral $M_{\text{tot}}$ posteriors and current EoS uncertainties, as performed in \citet{abbott_gw190425_2020} \emdashb in conjunction with the addition of a remnant-classification mixture model to our formalism and/or simultaneous inference of $\R$ and $M_{\text{thres}}$ using (e.g.) the relation between $M_{\text{thres}}$ and $\fpeak^{max}$ \citep{bauswein_revealing_2014}.

Finally, as for instance discussed in \citet{sarin_evolution_2021,bauswein_identifying_2019,chatziioannou_studying_2020} and mentioned in \S\ref{phase-transitions}, there is a possibility that during the merger or post-merger phase, the BNS remnant may experience a hadron-quark phase transition. As shown in \citet{bauswein_identifying_2019} and \citet{blacker_constraining_2020}, this phase transition can manifest as a mass-dependent offset in $\fpeak$ from what is predicted by the empirical relations discussed in this work, dependent on the strength of the phase transition. This phenomenon could enable us to extend our hierarchical formalism to additionally infer the presence/absence of a strong phase transition, and \emdashb if present \emdashb the transition strength and nature of the high-mass, post-transition regime of the EoS. Such an extension would lend significant utility to our analysis well into the era of next-generation detectors, even if unambiguous detections of post-merger signals become common.

% fin.

\section*{Acknowledgements}
The authors would like to thank James Clark and Tim Dietrich for many helpful conversations about BayesWave and multimessenger considerations, respectively. This research makes use of the Numpy \citep{harris_array_2020}, SciPy \citep{virtaneN_scipy_2020}, Pandas \citep{mckinney_data_2010}, Matplotlib \citep{hunter_matplotlib_2007} scientific computing packages, the BayesWave data analysis pipeline \citep{cornish_bayeswave_2015,littenberg_bayesian_2015,cornish_bayeswave_2021}, and utilized computing resources provided by the Minnesota Supercomputing Institute (MSI) at the University of Minnesota. The authors are grateful for computational resources provided by the LIGO Laboratory and supported by National Science Foundation Grants PHY-0757058 and PHY-0823459. This material is based upon work supported by NSF's LIGO Laboratory which is a major facility fully funded by the National Science Foundation. A.W.C. and J.M. acknowledge support by the National Science Foundation Research Traineeship Program `Data Science in Multimessenger Astrophysics' under grant No. 1922512. M.W.C. acknowledges support from the National Science Foundation with
grant numbers PHY-2010970 and OAC-2117997. A.B. acknowledges support by the European Research Council (ERC) under the European Union’s Horizon 2020 research and innovation programme under grant agreement No.\ 759253, and support by Deutsche Forschungsgemeinschaft (DFG, German Research Foundation) - Project-ID 279384907 - SFB 1245 and DFG - Project-ID 138713538 - SFB 881 (“The Milky Way System”, subproject A10) and support by the State of Hesse within the Cluster Project ELEMENTS. The work of T.S. is supported by the State of Hesse within the Cluster Project ELEMENTS, and the Klaus Tschira Foundation. T.S. is Fellow of the International Max Planck Research School for Astronomy and Cosmic Physics at the University of Heidelberg (IMPRS-HD) and acknowledges financial support from IMPRS-HD. T.S. acknowledges support by the High Performance and Cloud Computing Group at the Zentrum f{\"u}r Datenverarbeitung of the University of T{\"u}bingen, the state of Baden-W{\"u}rttemberg through bwHPC and the German Research Foundation (DFG) through grant no INST 37/935-1 FUGG. A.W.C. and J.M. contributed equally to this work.

\bibliography{HBPM_bib_122122_BB.bib}
\bibliographystyle{apsrev4-2.bst}

\end{document}